\documentclass[useAMS,usenatbib]{mn2e}

\usepackage{epsfig}
\newlength{\widthxxyy}
\title[Dust obscuration studies along quasar sight lines using simulated galaxies]{Dust obscuration studies along quasar sight lines using simulated galaxies}
\author[D. K. Churches and A. H. Nelson and M. G. Edmunds]{D. K. Churches\thanks{E-mail:
d.churches@astro.cf.ac.uk}, A. H.  Nelson and M. G. Edmunds
\\
Department of Physics and Astronomy, 5 The Parade, Cardiff, CF24 3YB}
\begin{document}

\date{Accepted 1988 December 15. Received 1988 December 14; in original form 1988 October 11}

\pagerange{\pageref{firstpage}--\pageref{lastpage}} \pubyear{2002}

\maketitle

\label{firstpage}

\begin{abstract}

We use the results of a set of three--dimensional SPH--Treecode 
simulations which model the formation and early evolution of disk galaxies, including the generation of heavy elements by star formation, to investigate the effects of dust absorption in quasar absorption line systems.

Using a simple prescription for the production of dust, we have compared the column density, zinc abundance and optical depth properties of our models to the known properties of Damped Lyman $\alpha$ systems.

We find that a significant fraction of our model galaxy disks have a higher column density than any observed DLA system. We are also able to show that such parts of the disk tend to be optically thick, implying that any background quasar would be obscured through much of the disk. This would produce the selection effect against the denser absorption systems thought to be present in observations.

\end{abstract}

\begin{keywords}
galaxies: abundances -- galaxies: formation -- quasars: absorption lines
\end{keywords}

\section{Introduction}



Damped Lyman Alpha (DLA) systems are clouds of gas with column densities greater than $2 \times 10^{20} \rm{HI cm^{-2}}$ which have been detected in QSO absorption line systems.

Their precise nature is still unknown, and they have been interpreted 
as the precursors of present--day massive disks (\cite{Prochaskaetal98}, \cite{Ferr97}), as
sub--galactic lumps still in the process of forming galaxies through hierarchical merging (\cite{Haehneltetal98}), and as low--surface brightness galaxies (\cite{Jimenezetal99}). 
\cite{Pett97} showed that there was a large spread in metallicity at any given redshift for DLAs, indicating that they were drawn from a population of galaxies of varying morphology and differing stages of chemical evolution.
\cite{Lu96} showed that iron abundances were 1/100 solar or less for systems at $z>3$, but for $z<3$ many had abundances ten times larger than this, implying that star formation had begun in earnest at about $z=3$.

An argument in favour of the disk precursor idea is the fact that the density of gas in the damped lyman alpha absorbers is similar to the density of matter in the form of stars in present--day galaxy disks. This number is approximately five times higher than the density of matter in the form of neutral hydrogen in the disks of present--day galaxies. 
The implication is that the gas in the damped Ly$\alpha$ systems at redshifts of $z \sim 3$ turns into stars, leaving the gas fraction in the $z=0$ disks at 10--20 \%. 

If the damped Ly$\alpha$ systems are massive disks at high redshift, then the profiles of metal absorption lines may be interpreted as being evidence for rotation (\cite{Prochaskaetal98}). If they are smaller clumps in the process of merging, the velocity profiles can also be explained by clump bulk motions (\cite{Haehneltetal98}). 

In recent years, numerical simulations have shown that DLAs can be produced naturally within the Cold Dark Matter picture of structure formation (\cite{Katzetal96}, \cite{Gardneretal97}, \cite{Gardneretal01}). \cite{Katzetal96} showed that in a standard cluster-normalised CDM model (SCDM), the number of DLA systems produced was close to observations, although the less--dense Lyman Limit (LL) systems were significantly under--produced. They found that the lines of sight which produced damped absorption typically passed near the centers of relatively dense, massive proto--galaxies with a size of $\sim$ 10 kpc.
\cite{Gardneretal97} used a high--resolution SPH simulation to calibrate the relation between the absorption cross section and circular rotation velocity of dark matter halos (the $(\alpha,v_c)$ relation), and then used this with the Press--Schecter formalism (\cite{PressSchecter74}) to calculate the number of DLAs per unit redshift in a number of different cosmological models. They showed that the observations could be fitted best by COBE--normalised CDM (CCDM) or SCDM, although it is well--known that CCDM produces excessively massive clusters at $z=0$, and that SCDM disagrees with the results of COBE.
\cite{Gardneretal01} carried out numerical simulations of 5 variants of the CDM cosmology. 
They found a clear relationship between the HI column density of each absorber and its projected distance to the nearest galaxy, with DLA absorption occuring for distances of less than 10-15 kpc, and LL absorption out to distances of 30 kpc. Considering only the structures resolvable in the simulation, all models underproduced both DLA and LL absorption abundances. In order to calculate the absorption from lower--mass halos, they calibrated the $(\alpha,v_c)$ relation using halos in the simulations and then extended down to lower masses using a mass function provided by \cite{Jenkinsetal01}. In doing this they show that all models are able to replicate observations of both DLA and LL absorption abundances so long as absorption is produced by halos down to $v_c \sim 30 - 50 \rm{km\,\,s^{-1}}$. The question of the effect of low--mass halos awaits numerical simulations which are able to resolve structures down to $v_c \sim 30 \rm{km\,\,s^{-1}}$.

In this work we too assume that DLAs are produced by the precursors of modern--day disk galaxies, but we re--investigate the possible importance of the presence of \emph{dust} in such galaxies. Dust can increase the optical depth of the galaxy disk, perhaps to the point where the background quasar becomes obscured.
We add to our analysis a simple model for the production of dust, and we calculate the optical depth through a set of galaxy disks which this produces.

The remainder of this paper is organized as follows.
In section \ref{code} we discuss the code and in section \ref{initconds} we discuss the initial conditions used. In sections \ref{numberdensities} and \ref{metallicities} we describe how we calculate number densities and metallicities along lines of sight. In section \ref{dustmodel} we introduce our dust model, and in section \ref{results} we discuss the results. Our conclusions are presented in section \ref{conclusions}.

\section{The model}
\subsection{The code}
\label{code}

The code which was used to perform the simulations is described in detail in \cite{Churchesetal01} and only a brief overview will be given here.

The code is a standard SPH--Treecode (\cite{BH86}, \cite{Ging77}, \cite{Lucy77}) which has been
augmented to include the effects of star formation and chemical evolution. The star formation rate per unit volume is given by a Schmidt Law (\cite{Schmidt59}), 
\begin{equation}
d \rho_{g}/dt = -k \rho_{g}^{n} \,\,,
\label{schmidt}
\end{equation}
where $\rho_{g}$ is the gas density and $k$ and $n$ are constants. Observations by Kennicutt (1998) suggest that $n \sim 1.5$.

Chemical evolution is modelled by an extension to the well--known \emph{simple model}, which takes a fixed
volume of the galaxy and turns gas into stars within it. This model makes the prediction that the abundance of heavy
elements within the volume varies with time via the following relation:

\begin{equation}
\frac{d(Zg)}{dt} = (p-Z) \alpha \frac{dS}{dt} \,\,,
\label{chem1}
\end{equation}
where $g$ is the total mass of gas within the box, a fraction $Z$ of which is in the form of heavy elements.
$dS$ is the mass of gas which has just been turned into stars in time interval $dt$, and the yield $p$ is the ratio of the mass of metals which is returned to the ISM to the mass which is lost to the stellar graveyard, per generation of
star formation.
The extension which we have made is to let the stars and gas decouple from each other and evolve as dynamically
separate systems, whereas in the simple model they forced to stay together in the closed box. This is achieved by
applying Eqs. (\ref{schmidt}) and (\ref{chem1}) to each SPH gas particle. We then form a population of collisionless stellar particles where those gas particles with a larger star formation rate are more likely to spawn a new stellar
particle, as is described fully in \cite{Churchesetal01}.

\subsection{The initial Conditions}
\label{initconds}
We follow the approach of \cite{Katz91}, \cite{KatzGunn91}, \cite{Katz92}, \cite{SM94}, \cite{Vedelatal94}, \cite{SteinMull95}, \cite{Williams01} and \cite{Churchesetal01} which is to model the collapse of
an isolated spherical perturbation. These models have a resolution which is less than 1~kpc, and so can be used to probe internal structure, and have had great success in forming realistic disk galaxies. 

We start with a spherical dark matter halo which is composed of 6000 collisionless particles, and which is in solid body rotation (see below). Embedded in this are 6000 SPH gas particles which will model the gaseous component. As the simulation procceds, gas is converted into stars, and collisionless star particles are formed. 
 
The effect of tidal torques by neighbouring halos on the dark halo to be modelled is approximated by setting the system in solid body rotation. The amount of rotation is characterised by the dimensionless spin parameter $\lambda$, which is given by
\begin{equation}
\lambda = J \,|E|^{1/2} \, G^{-1} \, M^{-5/2} \,\,,
\label{spin1}
\end{equation}
where $J$ is the total angular momentum, $E$ the total energy, $M$ the total mass of the system and $G$ the gravitational constant. The theoretical prediction of Peebles (1969) is that $\overline{\lambda} \sim 0.08$, a value which has been confirmed by N--body experiments. 

The papers of \cite{KatzGunn91}, \cite{Katz92}, and \cite{SteinMull95} add small--scale noise to their initial conditions in an attempt to model the expected CDM power spectrum. However, it was shown by \cite{KatzGunn91} that the final properties of the galaxies were essentially independent of the amount of small--scale noise. For this reason, we choose to describe the small--scale noise in the system at the start of the simulation using simple Poisson noise. Experiments have been performed to verify
that all of the physical properties of our numerical galaxies are independent of the particular realisation of the noise.

The remaining parameters which specify the system are $k$ and $n$ in the Schmidt Law. Analytical work by \cite{EG95} has shown that in order for the heavy element abundance to decrease with radius, $n$ must be greater than unity, while observations described by \cite{Kenn98} indicate that $n \sim 1.5$, which is the value we have used.

If one makes the assumption that a gas disk with a mass of $M=5\times 10^{10} \rm{M_{\odot}}$, a radius of 10~kpc and height of 200~pc turns 90\% of its mass into stars over a timescale of 5~Gyr with $n=1.5$, then the value for $k$ thus obtained is 

\begin{equation}
k = 1.45 \,\,  \rm{M^{-1/2}_{\odot} \, pc^{3/2} \, Gyr^{-1}}
\label{kest1}
\end{equation}
which we use here. 

A much wider range of initial conditions were investigated in \cite{Churchesetal01}, \cite{Churches99} and \cite{Williams01}.  
These simulations have shown that the parameters which have the most significant effect upon the final properties of the model galaxies are the total mass of the system and its angular momentum. Therefore here we present results for three different values of each. We choose three values for the total mass: $M=2.5\times 10^{11} \rm{M_{\odot}}$,$M=5\times 10^{11} \rm{M_{\odot}}$ and $M=1\times 10^{12} \rm{M_{\odot}}$. For one of these masses ($M=5\times 10^{11} \rm{M_{\odot}}$), we varied the initial angular momentum, producing the following values of the dimensionless spin parameter: $\lambda=0.06$, $\lambda=0.09$ and $\lambda=0.12$.

The simulation parameters are shown in table \ref{paramstable}. Runs 1,2 and 3 investigate the effect of varying $\lambda$ while keeping $M$ fixed, and runs 2, 4 and 5 investigate the effect of varying $M$ for a fixed $\lambda$.
\begin{table}
\begin{center}
\begin{tabular}{|c|c|c|} \hline
 run      &  total mass $M(\times 10^{11} \rm{M_{\odot}})$     & Spin Parameter $\lambda$  \\ \hline
  1       &      5                                           &   0.06                    \\ 
  2       &      5                                           &   0.09                    \\ 
  3       &      5                                           &   0.12                    \\ 
  4       &      2.5                                         &   0.09                    \\ 
  5       &      10                                          &   0.09                    \\ \hline
 
\end{tabular}
\caption[]{The simulation parameters. We varied the total mass $M$ and the dimensionless spin parameter $\lambda$.}
\label{paramstable}
\end{center}
\end{table}

\

\

\subsection{Calculation of number densities}
\label{numberdensities}
The surface density is defined as the mass per unit area of a galaxy.
If one knows the mass of each of the particles in the galaxy, 
it is trivial to convert the surface density into a column density,
which is the number of particles per unit area. 

Numerically we use SPH to calculate a smoothed estimate of the surface density throughout our simulated galaxies and then we convert this into a column density of hydrogen, using the relation $10^{21} \rm{HI cm^{-2}}=8 \, \rm{M_{\odot}\,pc^{-2}}$.


 
 
 
For each of our models, we choose a particular time in its evolution. We then orient it so that it is face--on in the $(x,y)$ plane, and we lay down a square grid of equally spaced 
points which cover the galaxy. 
The grid is a square of side 30~kpc, centered on the galaxy. There are 
$16 \times 16 = 256$ grid points altogether, each separated by a distance of 2~kpc.

We then evaluate the column density at each sample point, and from this we can construct the distribution of column densities for that galaxy. 
This procedure is then repeated for each galaxy at different times and for
different inclination angles.

In real systems, the QSO may be regarded as a point source and the sight lines are very small in width. However, the sight lines pass through at least the thickness of the galaxy disk so that the absorption is averaged over a linear region in the ISM of at least a few hundred parsecs. 
This is comparable to the resolution of our simulations, so our column density calculations are sufficiently resolved.
 
\subsection{Calculation of metallicites}
\label{metallicities}
Also of interest to us is the gas metallicity $Z$ along any line of sight,
which is defined as the ratio of the total mass of metals along the column 
to the total mass of gas along the column. 
Given that in the simulations we know both the mass $m_{i}$ and metallicity 
$Z_{i}$ of each SPH gas particle (and therefore the mass of metals associated 
with each gas particle is $m_{i} Z_{i}$), this is a simple calculation.
 
We need to compare our results to observations of real damped
Lyman alpha systems. However, instead of $Z$, observers usually measure 
the abundance of a particular heavy element $X$ with respect to hydrogen,  $\left[X/H\right]$. 

Often $[Zn/H]$ is used as an abundance indicator because zinc is not believed to become significantly
depleted onto dust grains. Therefore a measurement of $[Zn/H]$ for the gas is more
likely to represent the true zinc abundance in the object.
Pagel et al.\ (1992) give an estimate for the relationship between 
the overall heavy element abundance $Z$ and $O/H$, $Z \sim 23 (O/H)$. 
Using this, we can express the abundance of $Zn$ 
with respect to
hydrogen, under the assumption that all the heavy elements increase in step with one
another. The result is (\cite{Churches99})

\begin{equation}
\log_{10} Z = \left[ \frac{Zn}{H} \right] - 1.77 \,\,.
\label{logz6}
\end{equation}

\subsection{A simple dust model}
\label{dustmodel}
We use an elementary model for the absorption properties of dust in
galaxies. This model is presented in \cite{Mathlinetal01} and \cite{Churches99}, and so only a summary of the main result is given here.

Our model calculates the optical depth $\tau$ which is produced by looking through a column of gas which has column density $N_{gas}$ and metallicity $Z$. We assume that a constant fraction $\eta$ of the heavy elements are in the form of dust, an assumption for which plausible physical arguments can be made (\cite{Edmunds01}). The dust is uniformly mixed with the gas and is composed of dust particles with radius $r_{dust}$ and density $\rho_{dust}$. This gives 

\begin{equation}
\tau = \frac{3 \eta Z 1.3 \rm{m_{H}} N_{gas}}{4 r_{dust} \rho_{dust}} \,\,.
\end{equation}

For $r_{dust}$ and $\rho_{dust}$ we have taken $r_{dust}=1 \times 10^{-7}$~m and $\rho_{dust}=2000 \,\rm{kg \,\,m^{-3}}$, and for $\eta$, \cite{EE98} suggest that $\eta=0.5$ is acceptable.
So
\begin{equation}
\tau = 4.07 \times 10^{-20} \,\rm{cm^{2}}\,\, Z \,\, N_{gas} \,(\rm{cm^{-2}})
\label{tau5}
\end{equation}

Using equation \ref{tau5} we can put an estimate on the optical depth produced by
looking through a region of a galaxy if it has metallicity $Z$ and column density
$N_{gas}$. This equation may be expressed in terms of $[Zn/H]$ instead of $Z$ using
equation \ref{logz6}.

\section{Results}
\label{results}

The simulations presented here are drawn from a set previously described in \cite{Churchesetal01} and a detailed description of their evolution is given there. However, a short recap will be given here.

The systems start off as uniform density spheres of gas and dark matter with a radius of 100 kpc. They are rotating about the $z$--axis, and are about to collapse. Collapse occurs approximately on a free--fall timescale, which is of the order of 1 Gyr.
Material from above and below the $x-y$ plane collide, the gas particles do not interpenetrate and so a flattened disk is formed. 
Rotational support about the $z$--axis gives the disk a radius of $\sim 15$~kpc. Knots and arm--like structures are formed, which subsequently merge as the system settles down.
By this time, because of the high gas densities, approximately 40\% of the initial gas mass has been turned into stars. 

As a natural result of the collapse, the gas assumes an exponential surface density profile, and any subsequent star formation occurs in an exponential disk. In \cite{Churchesetal01} we showed that these models produce galaxies which have negative abundance gradients in the range $(-0.02 \rightarrow -0.07)\,\,dex/kpc$, which is in broad agreement with observations.

We present snapshots of the absorption properties of our models at times of 1, 2, 3 and 5~Gyr. To give a simple indication of the approximate redshift reached for each of these times, we have used the equations for the Einstein--de--Sitter solution with $\Omega=1$ and $\rm{H_{0}}=75 \rm{km s^{-1} Mpc^{-1}}$. 
Table \ref{zoft} shows how, in this case, the time elapsed in the simulation is related to the redshift reached. Observations typically show the majority of DLAs lying in the range $1 < z < 4$, so the overlap of the models with observations is primarily for the first 2 Gyr of each model.

\begin{figure}
\begin{center}
\psfig{file=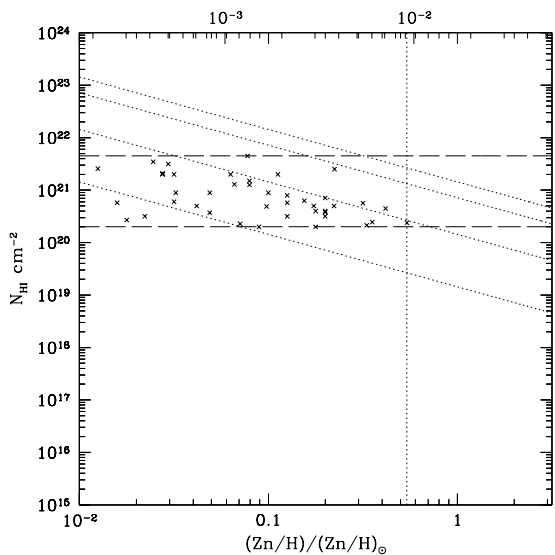,width=80mm,clip=}
\caption[Column density against zinc abundance for observed damped Lyman alpha
 systems.]
{Column density against zinc abundance for observed damped Lyman alpha
systems. 
References to the sources of data given in
Mathlin~\emph{et~al}~2001. See text for details of the dotted and dashed lines.}
\label{dla_obs}
\end{center}
\end{figure}

For each snapshot we lay down a grid which covers the gaseous extent of the galaxy and then we pierce the galaxy with 256 sight lines. For each sight line we calculate the column density and the zinc abundance, and from these two quantities we then calculate the optical depth using equation \ref{tau5}.

Figure \ref{dla_obs} shows the column density--abundance plot for the observed DLA systems. 
Running along the top of figure \ref{dla_obs} is the gas metallicity $Z$, where solar
composition is $Z \sim 0.02$. 
The upper horizontal line shows the position of a column density of $4.5 \times
10^{21}\,\,\rm{HI cm^{-2}}$, the column density of the densest damped Lyman alpha
system observed. The lower horizontal line corresponds to a column density
of $2 \times 10^{20}\,\,\rm{HI cm^{-2}}$, which is the lower limit for 
classification as a damped Lyman alpha system. 
The vertical dotted
line shows the highest zinc abundance so far observed in a damped Lyman alpha system,
that of $(Zn/H)/(Zn/H)_{\odot} \sim 0.54$.
The sloping dotted curves show column
density against abundance for constant optical depth values as
expected from the simple dust cloud model. From the bottom upwards
they correspond to optical depths of
0.01, 0.1, 0.5 and 1.0 respectively. 

One immediately notices from this diagram that no systems lie above the $\tau=0.5$ line. Therefore, interpreted in terms of our dust model, the observations all have optical depths less than $\tau=0.5$.

\begin{figure}
\begin{center}
\begin{tabular}{@{}lr@{}}
\psfig{file=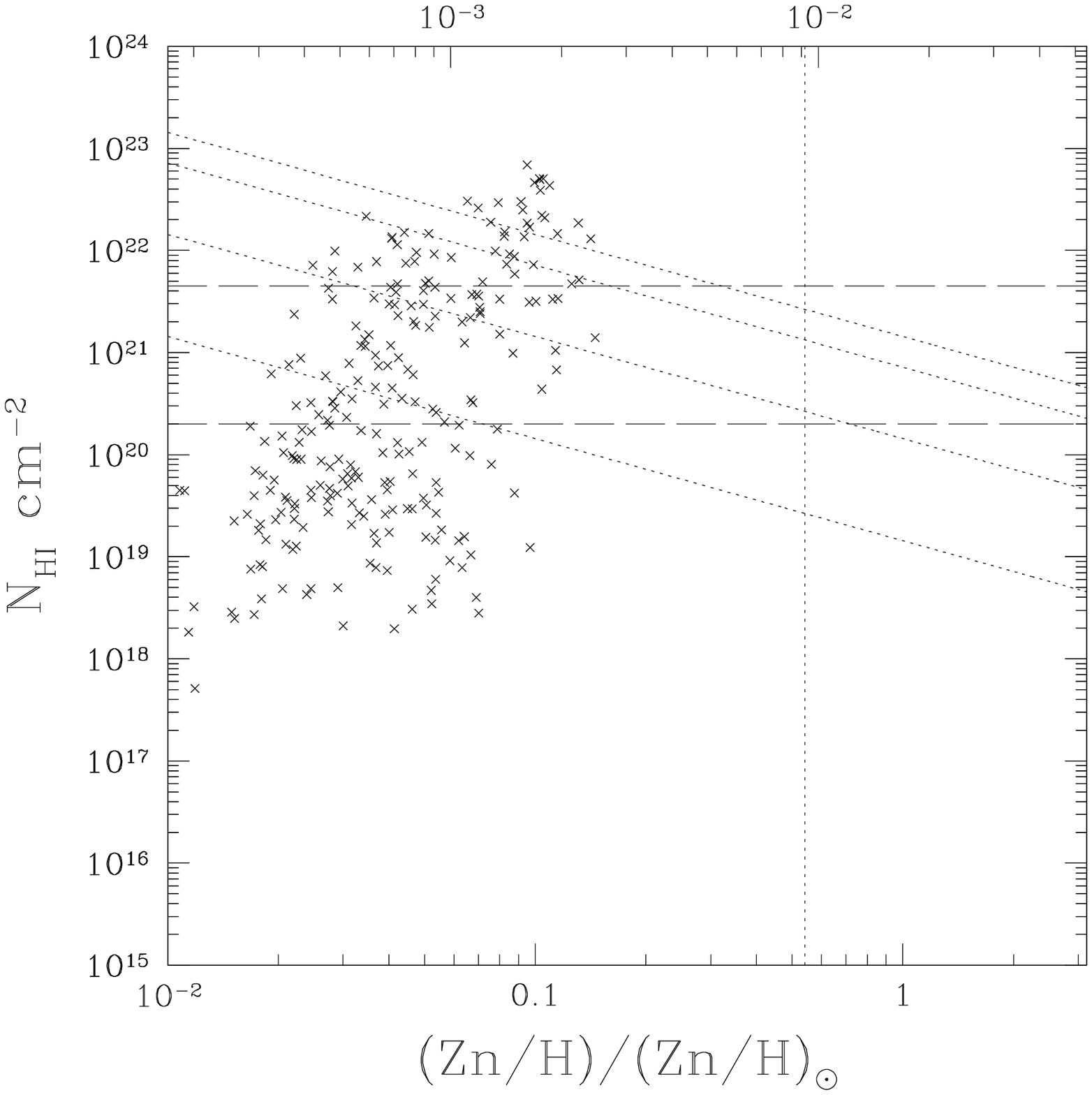,width=\widthxxyy,clip=}
&
\psfig{file=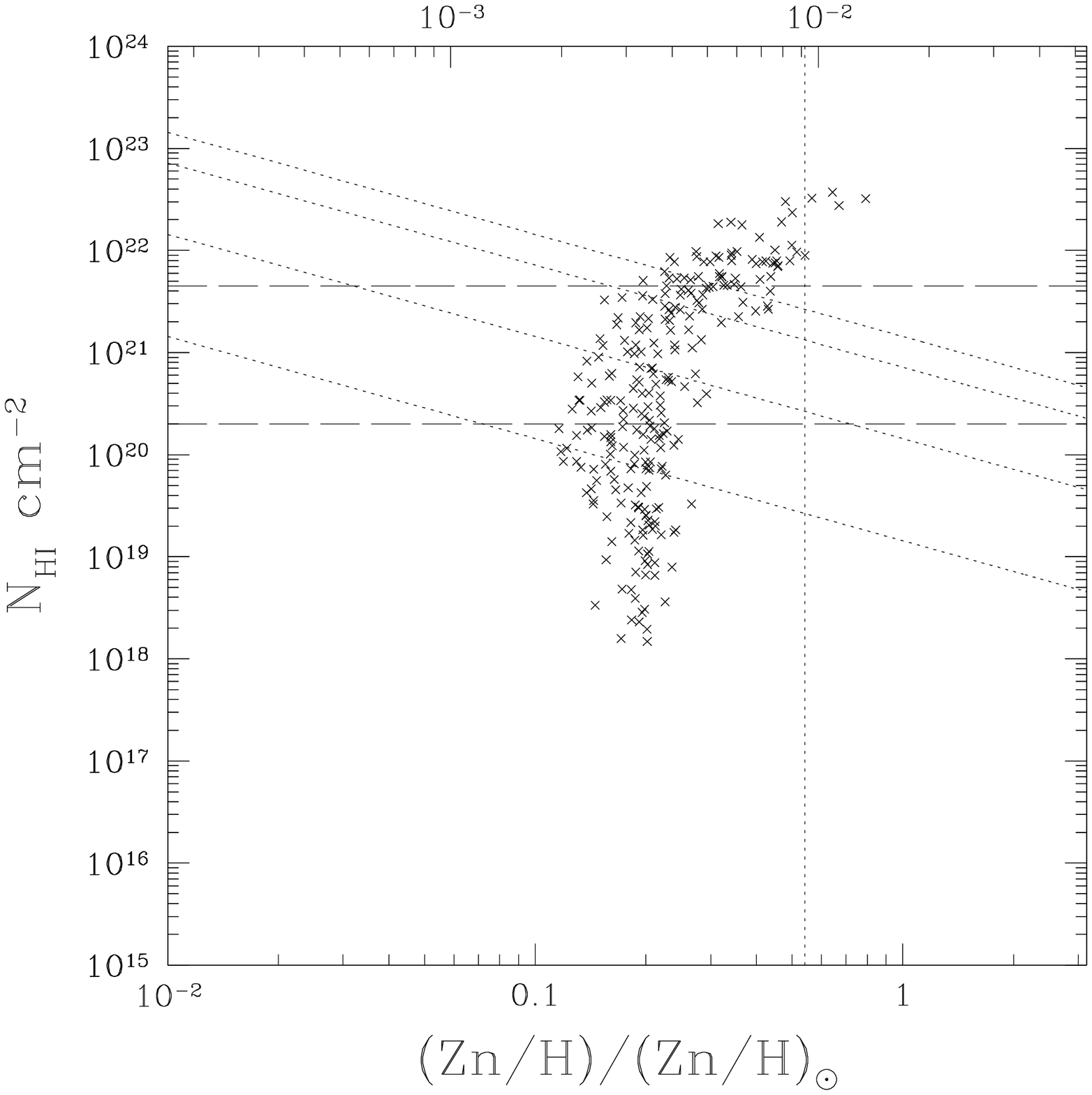,width=\widthxxyy,clip=}
\\
\psfig{file=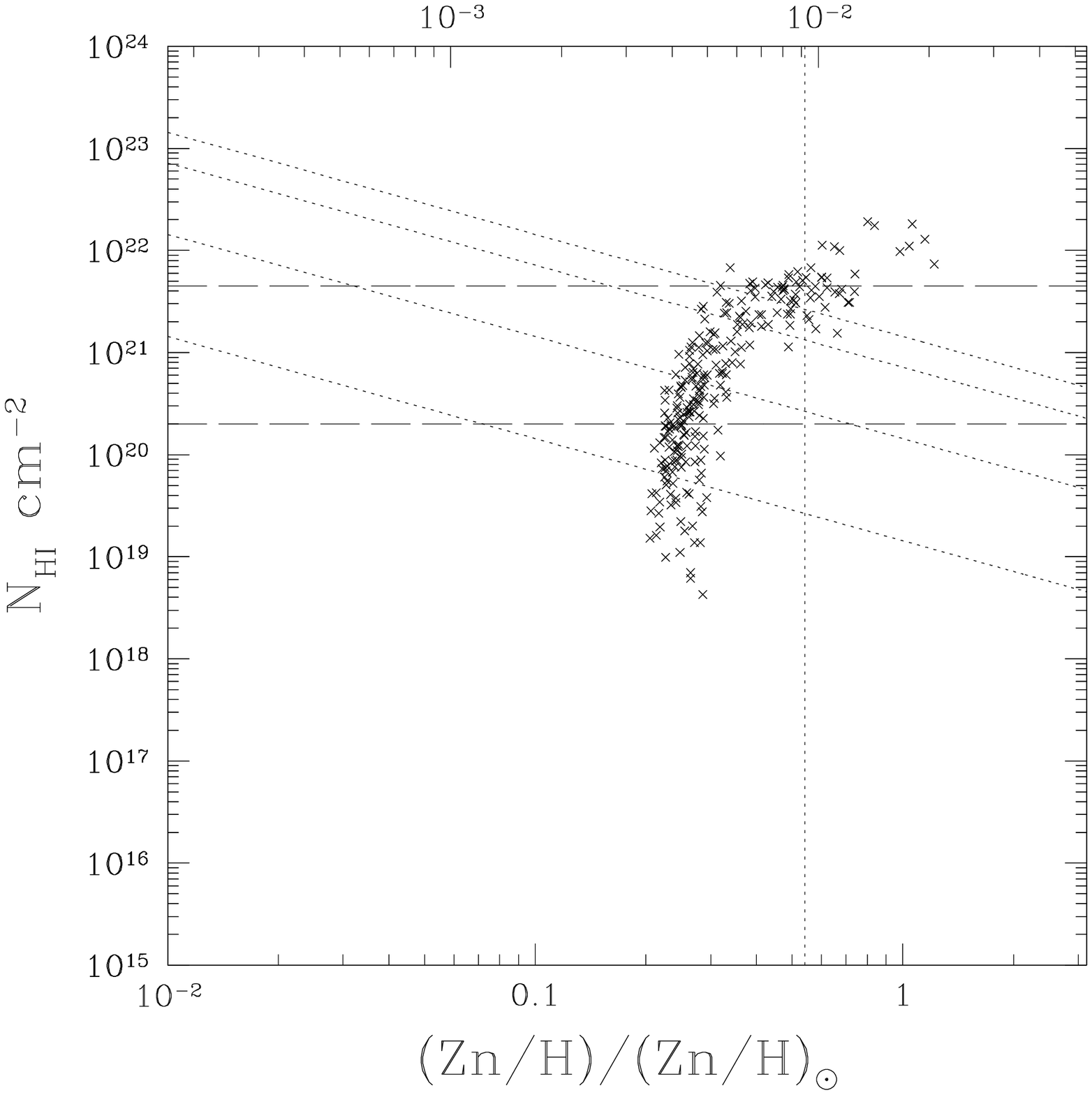,width=\widthxxyy,clip=}
&
\psfig{file=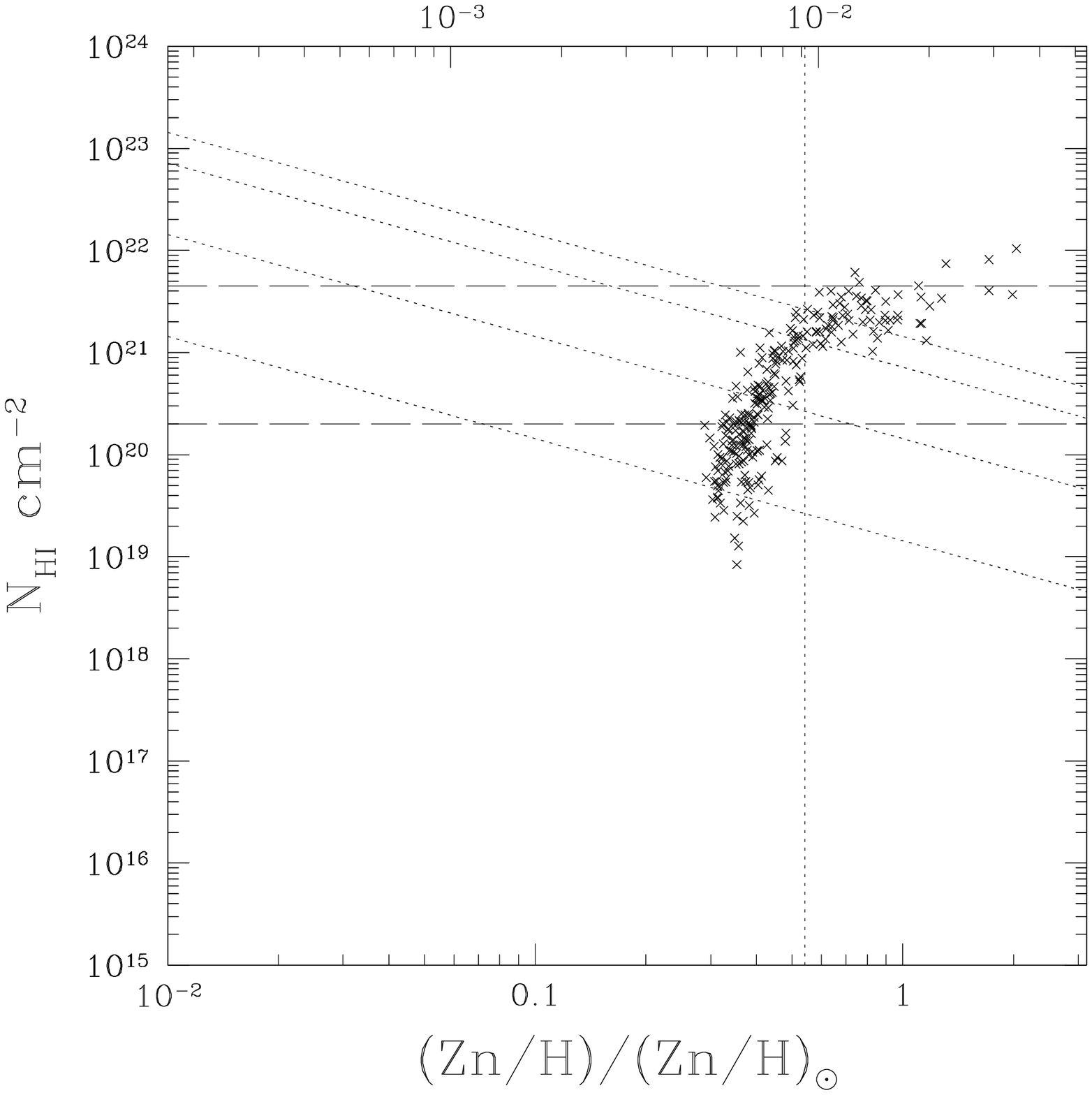,width=\widthxxyy,clip=}
\end{tabular}
\caption[]{Column densities and zinc abundances of
lines of sight through a model galaxy with mass
$M=5\times 10^{11} \rm{M_{\odot}}$ and $\lambda=0.06$. See text for a description of this figure.} 
\label{run1data}
\end{center}
\end{figure}

\begin{figure}
\begin{center}
\begin{tabular}{@{}lr@{}}
\psfig{file=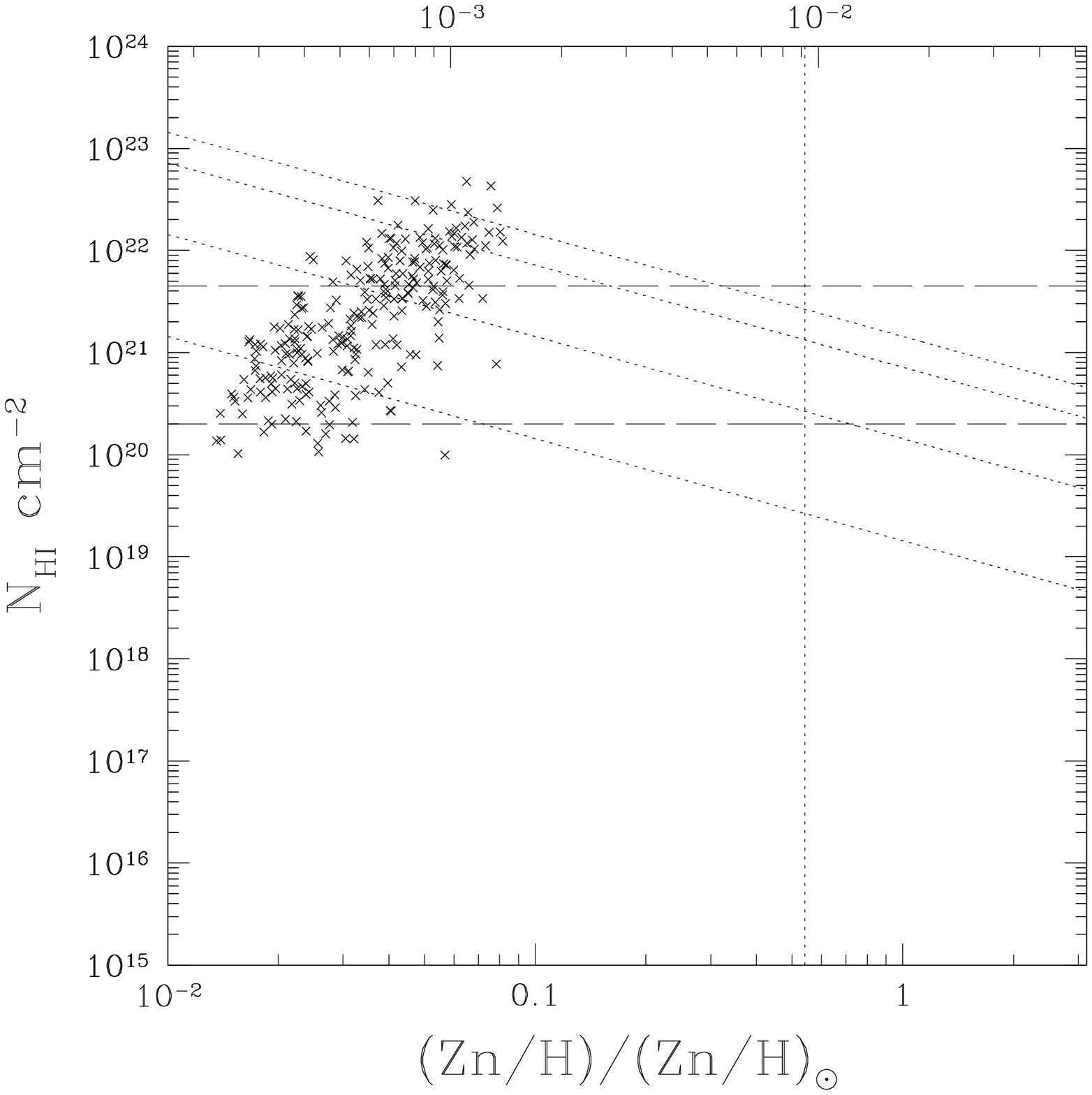,width=\widthxxyy,clip=}
&
\psfig{file=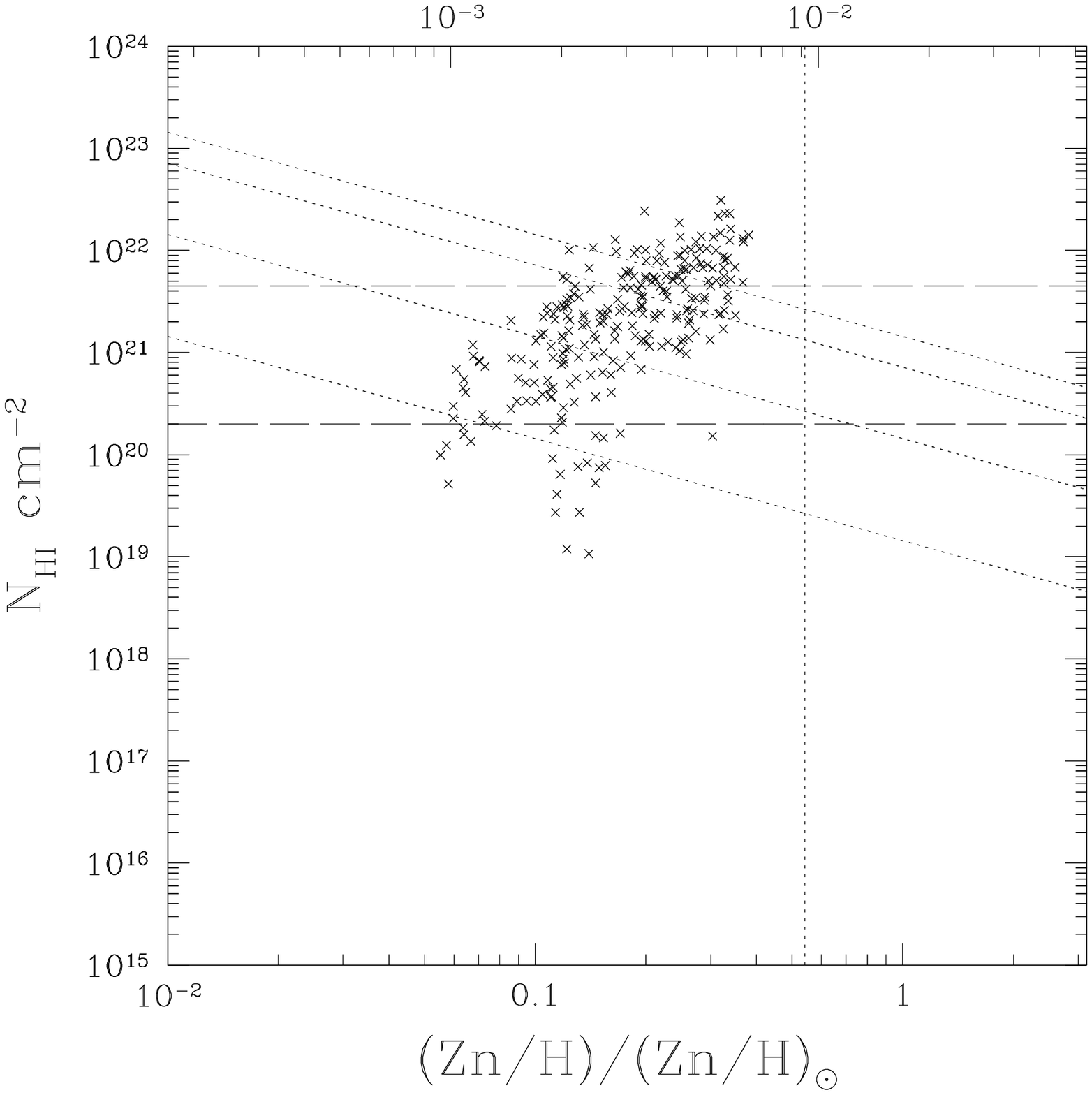,width=\widthxxyy,clip=}
\\
\psfig{file=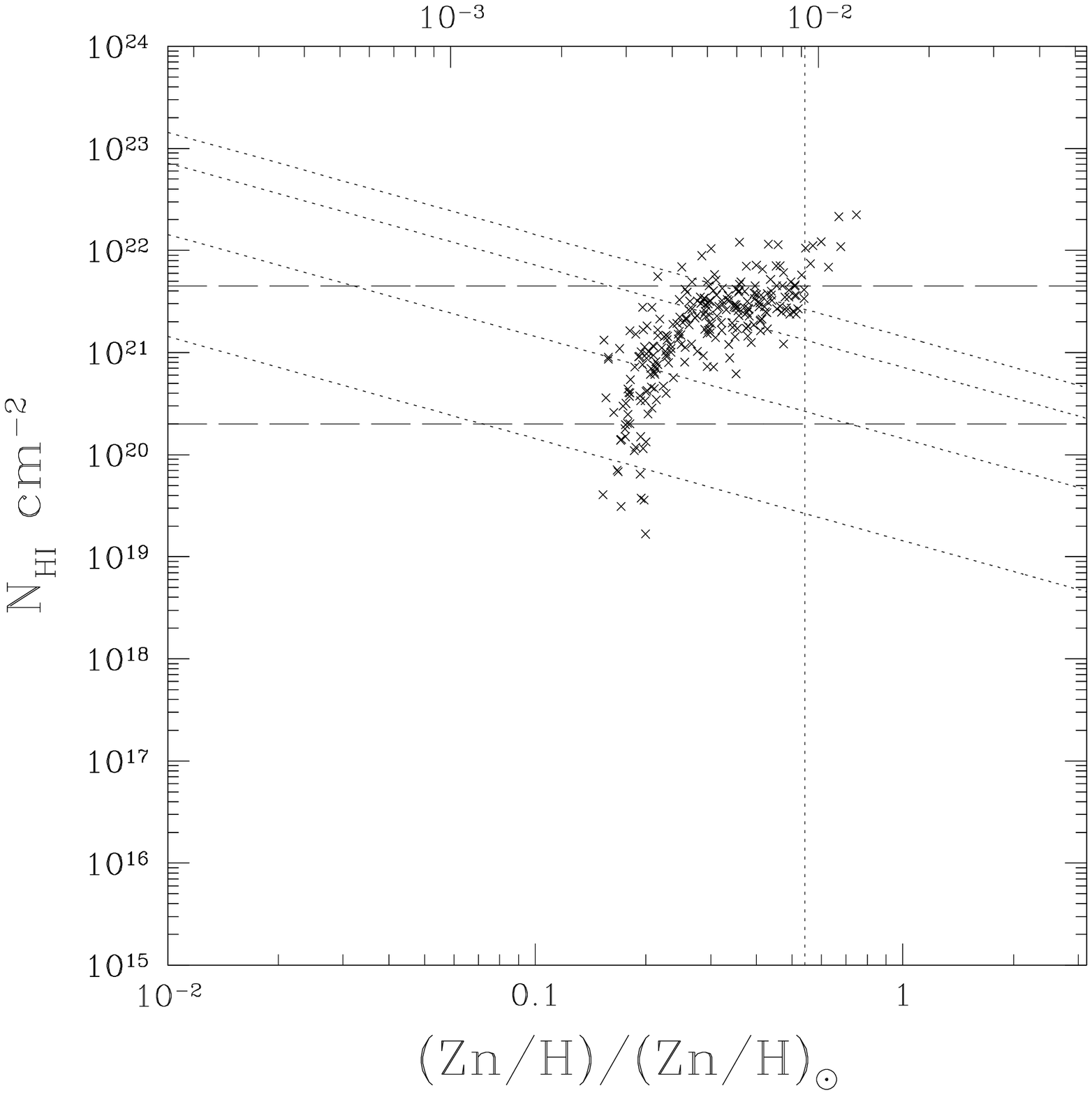,width=\widthxxyy,clip=}
&
\psfig{file=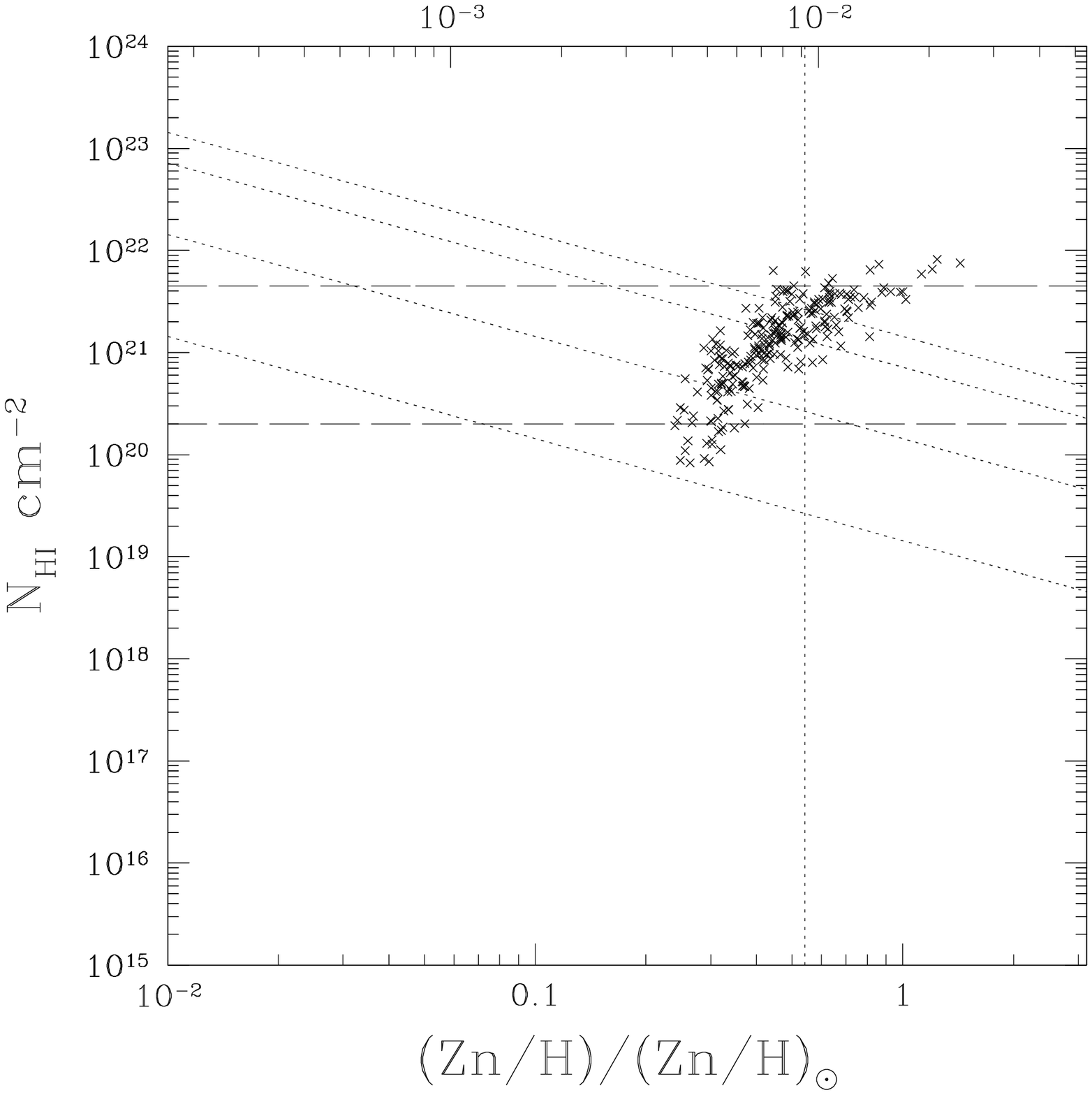,width=\widthxxyy,clip=}
\end{tabular}
\caption[]{Column densities and zinc abundances of
lines of sight through a model galaxy with mass
$M=5\times 10^{11} \rm{M_{\odot}}$ and $\lambda=0.09$.}
\label{run2data}
\end{center}
\end{figure}

\begin{figure}
\begin{center}
\begin{tabular}{@{}lr@{}}
\psfig{file=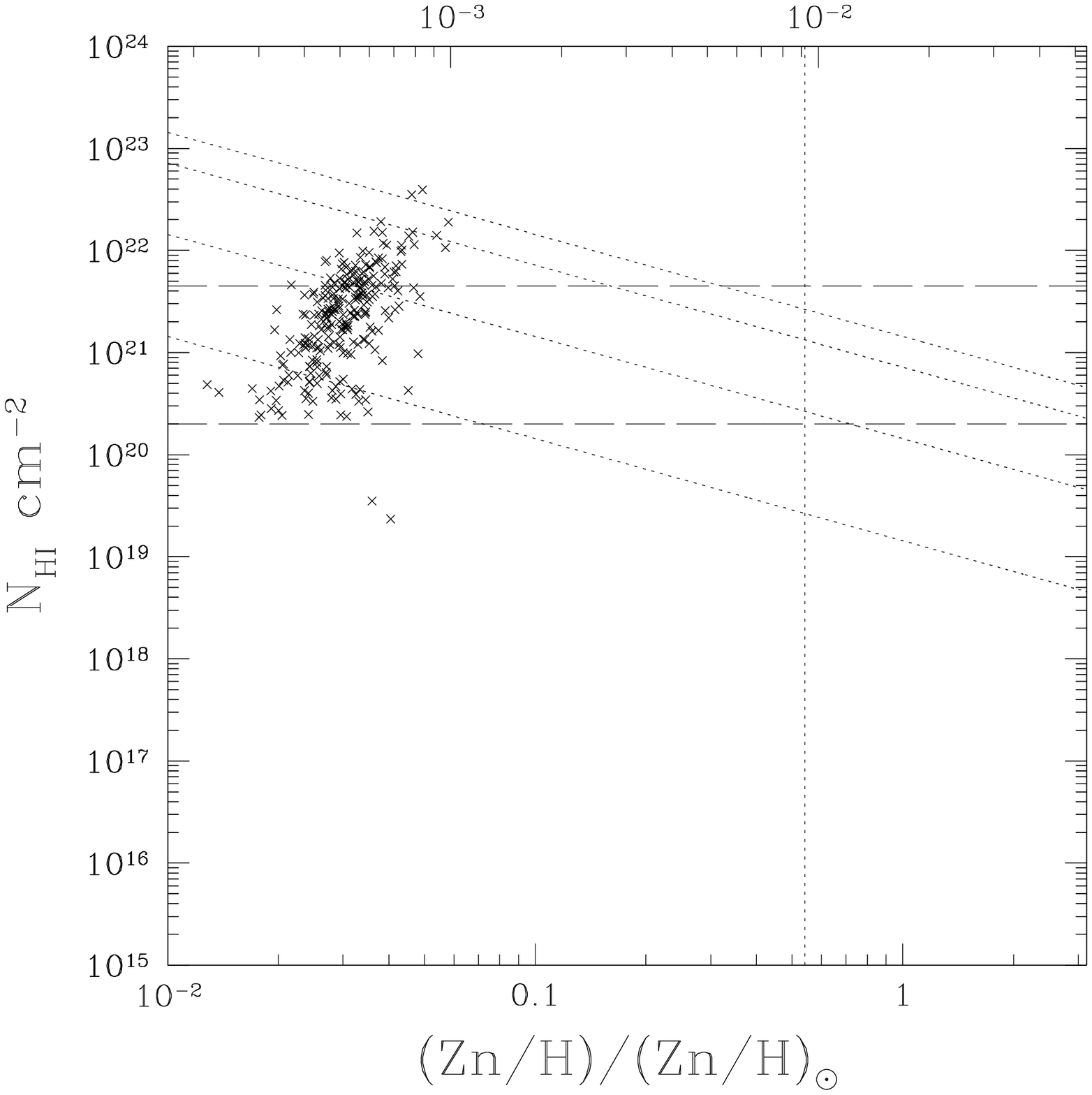,width=\widthxxyy,clip=}
&
\psfig{file=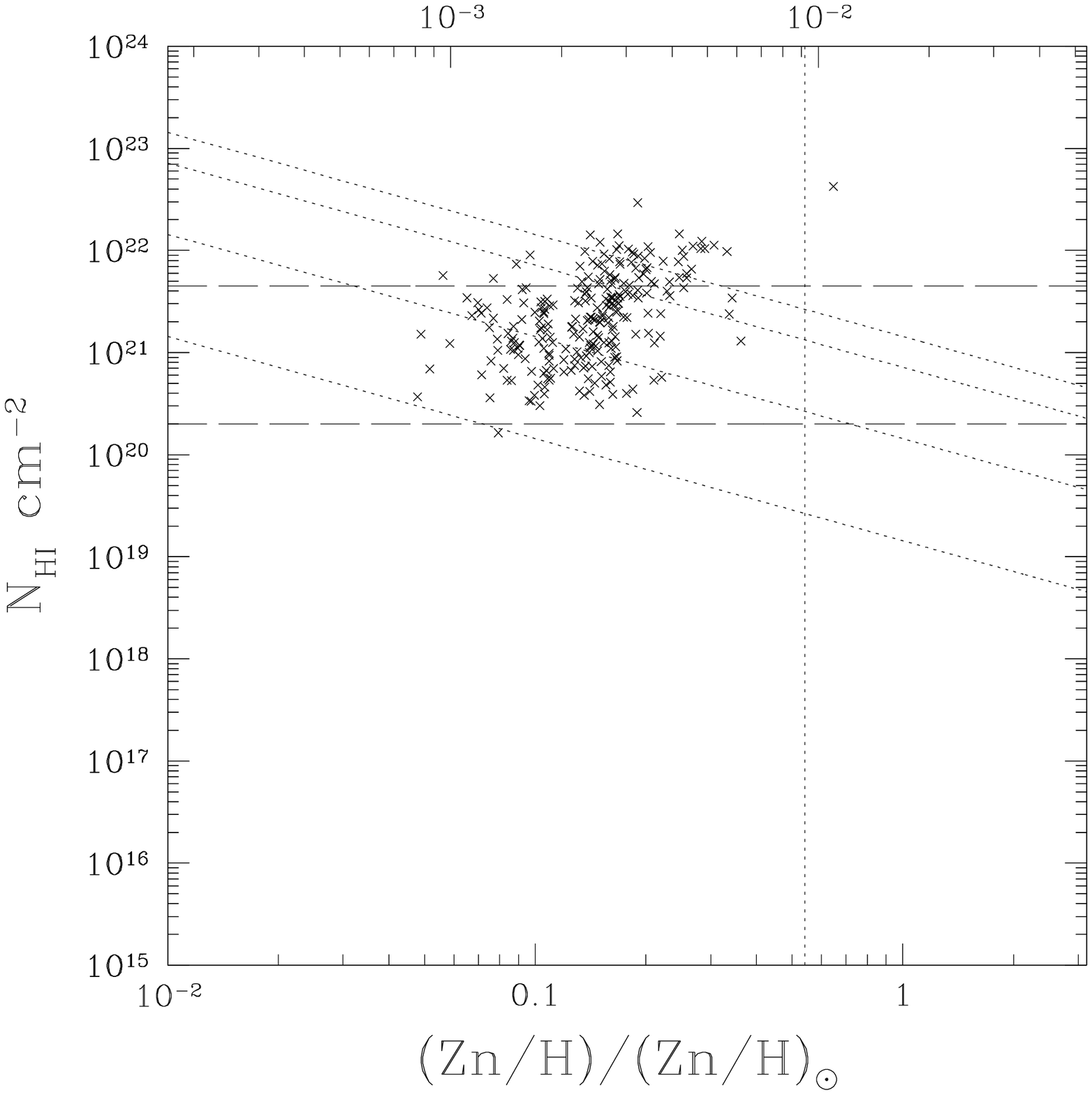,width=\widthxxyy,clip=}
\\
\psfig{file=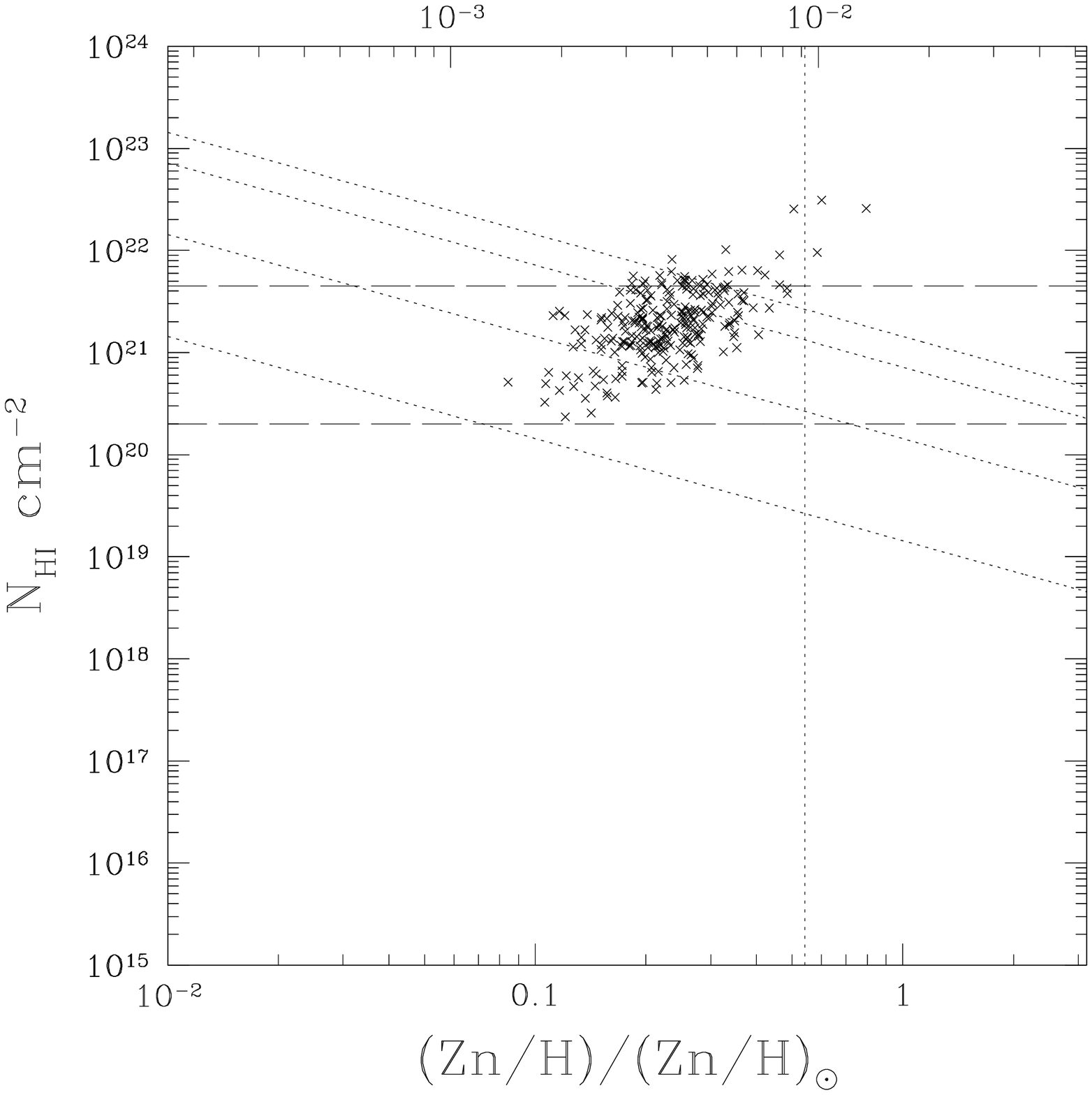,width=\widthxxyy,clip=}
&
\psfig{file=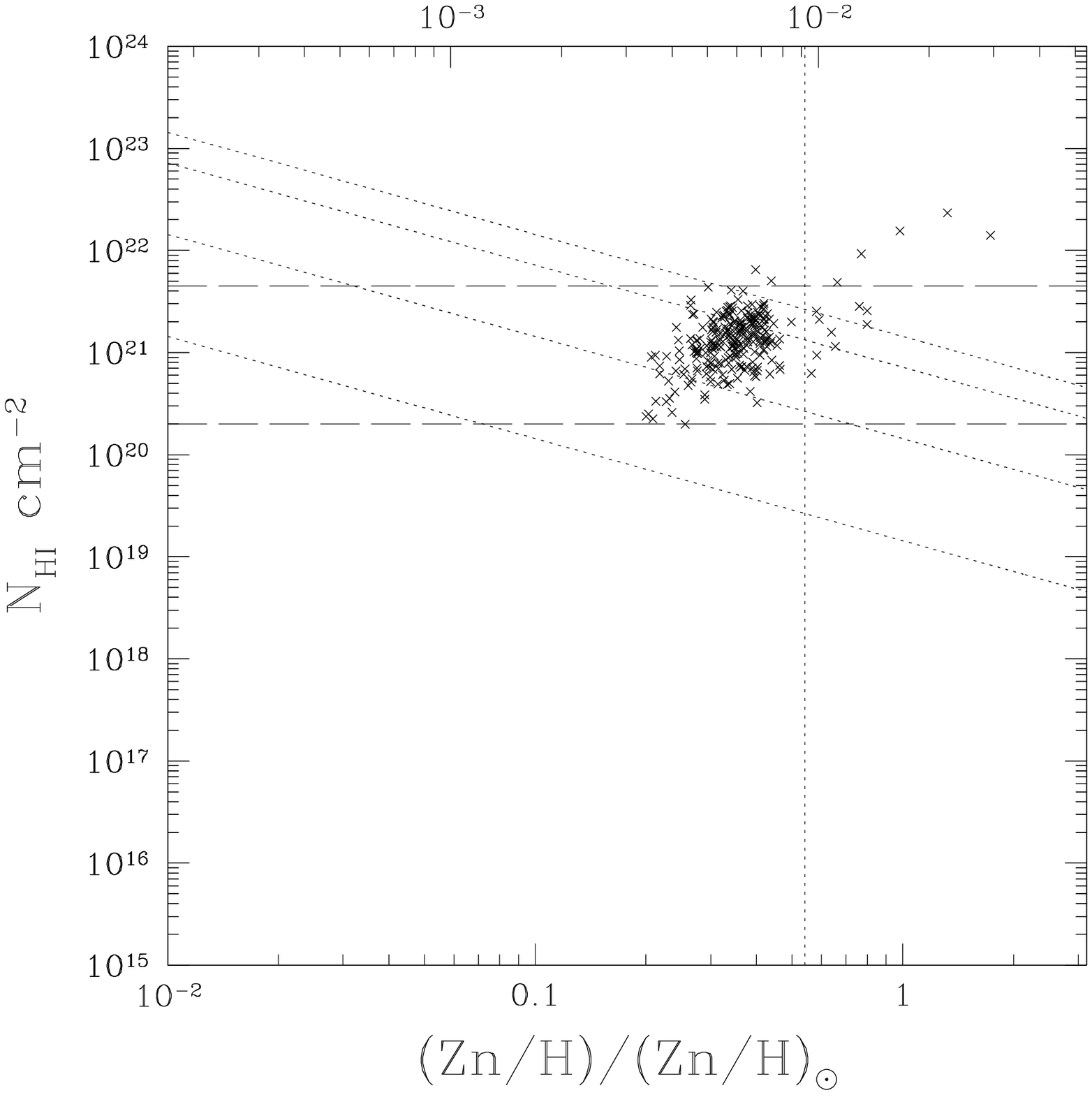,width=\widthxxyy,clip=}
\end{tabular}
\caption[]{Column densities and zinc abundances of
lines of sight through a model galaxy with mass
$M=5\times 10^{11} \rm{M_{\odot}}$ and $\lambda=0.12$.}
\label{run3data}
\end{center}
\end{figure}

\begin{figure}
\begin{center}
\begin{tabular}{@{}lr@{}}
\psfig{file=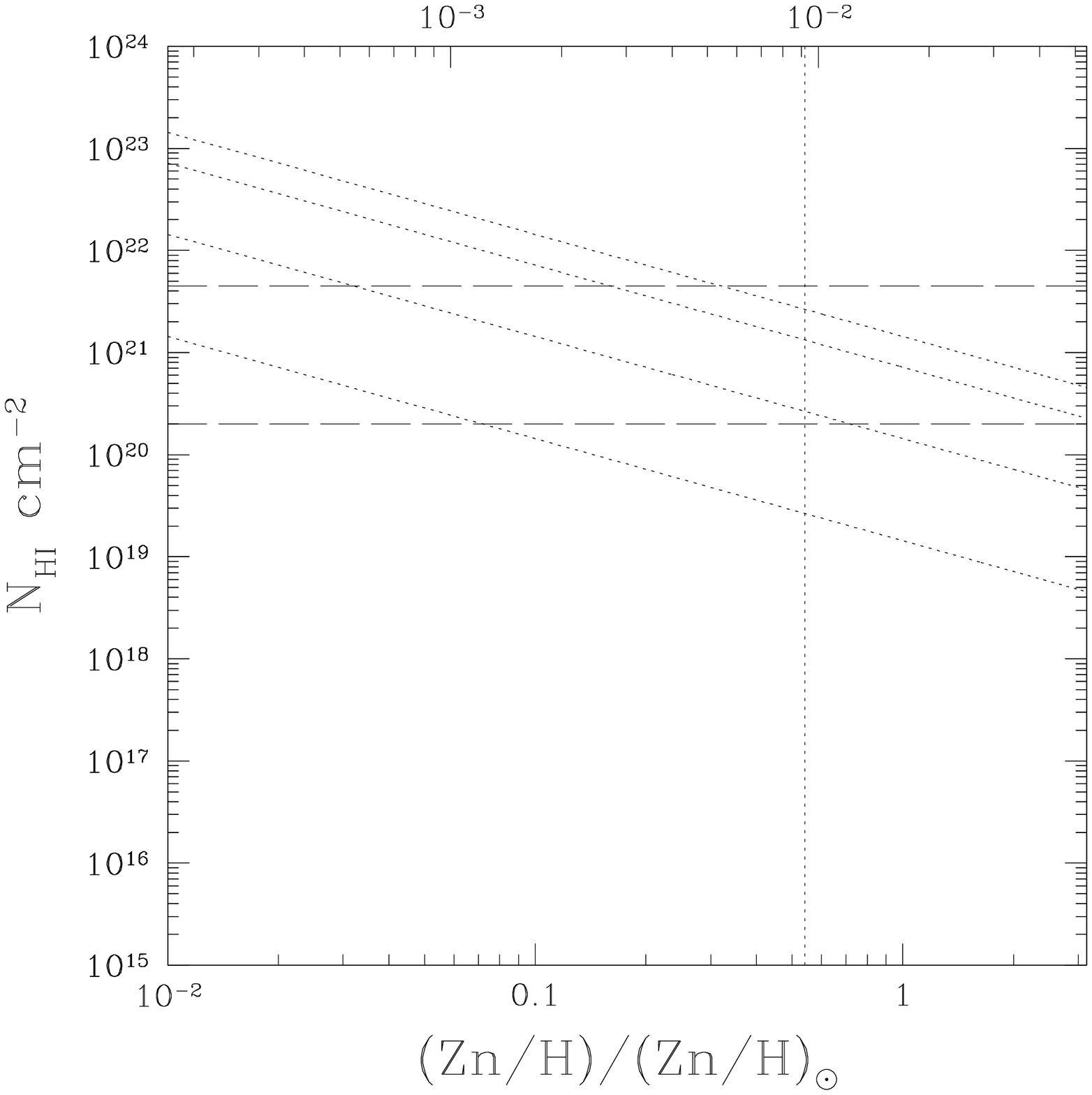,width=\widthxxyy,clip=}
&
\psfig{file=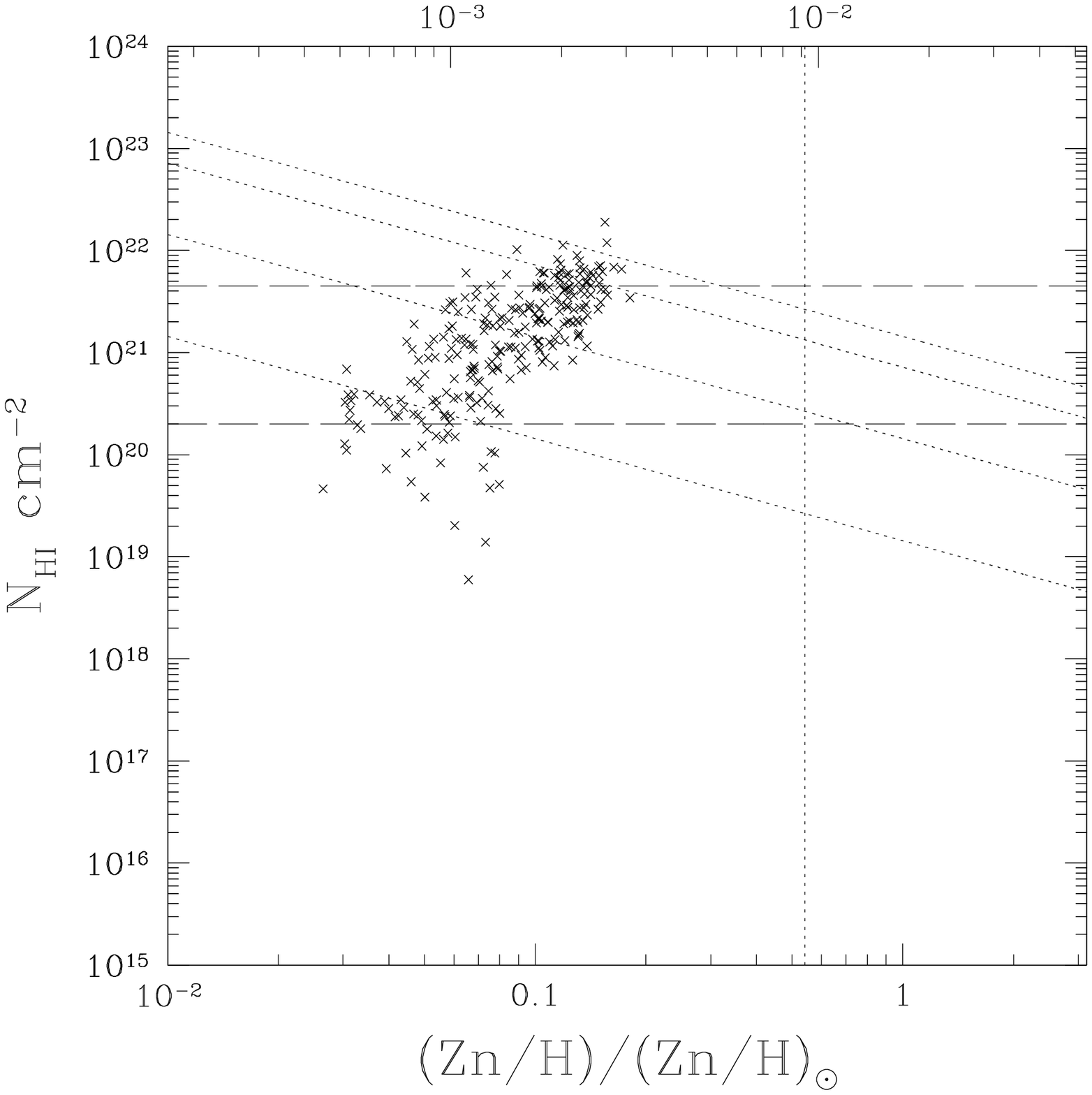,width=\widthxxyy,clip=}
\\
\psfig{file=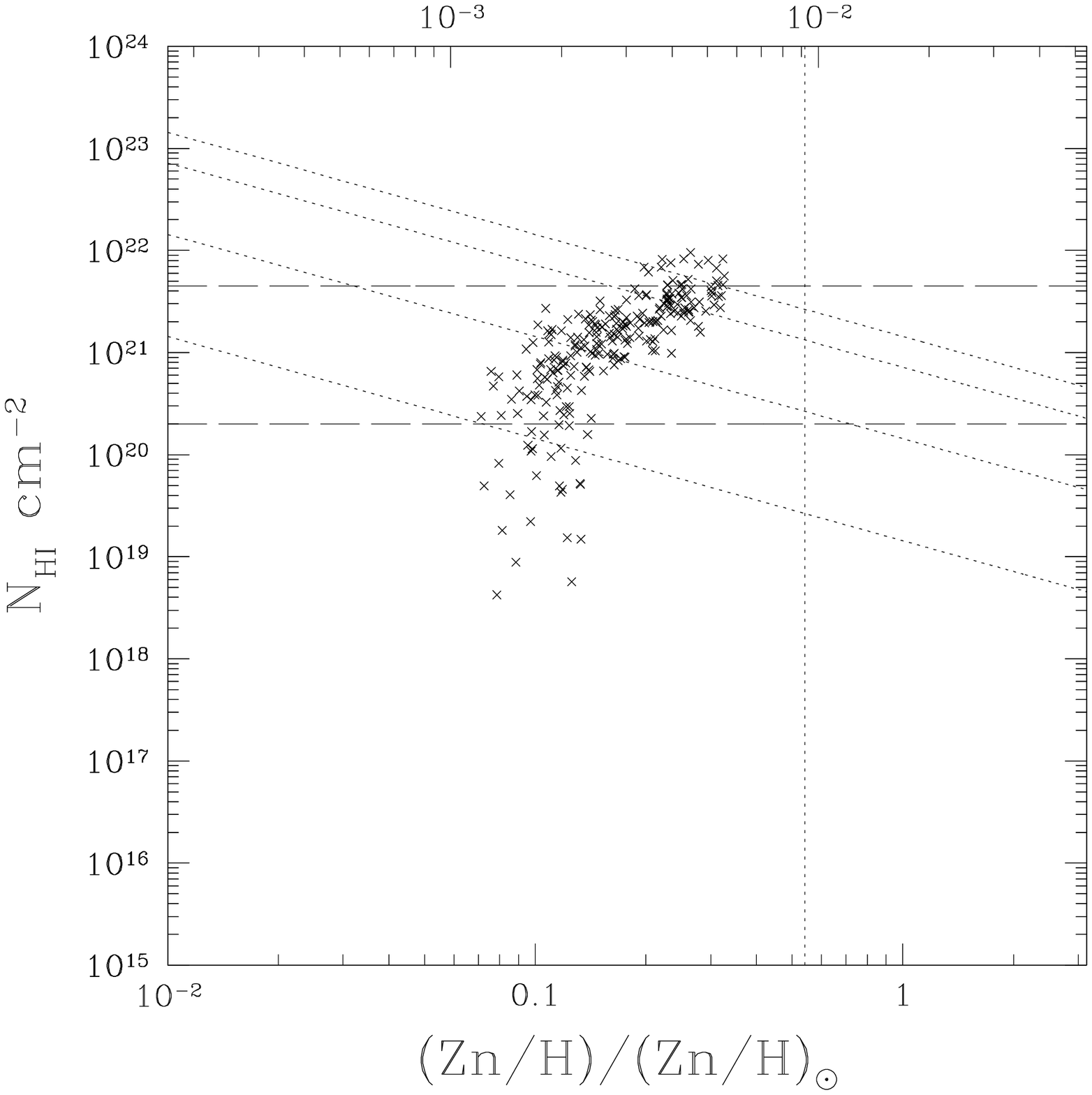,width=\widthxxyy,clip=}
&
\psfig{file=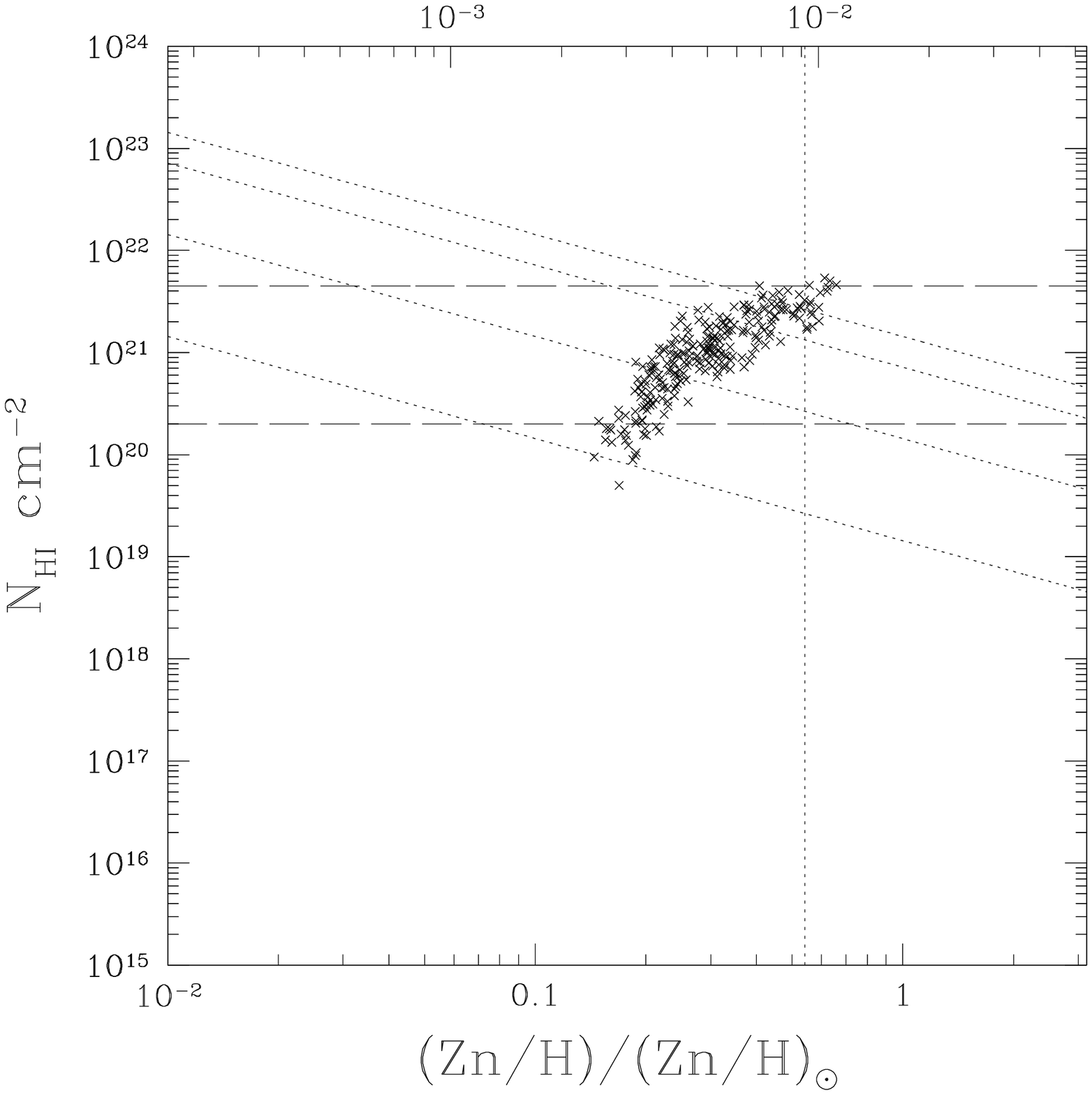,width=\widthxxyy,clip=}
\end{tabular}
\caption[]{Column densities and zinc abundances of
lines of sight through a model galaxy with mass
$M=2.5\times 10^{11} \rm{M_{\odot}}$ and $\lambda=0.09$.}
\label{run4data}
\end{center}
\end{figure}

\begin{figure}
\begin{center}
\begin{tabular}{@{}lr@{}}
\psfig{file=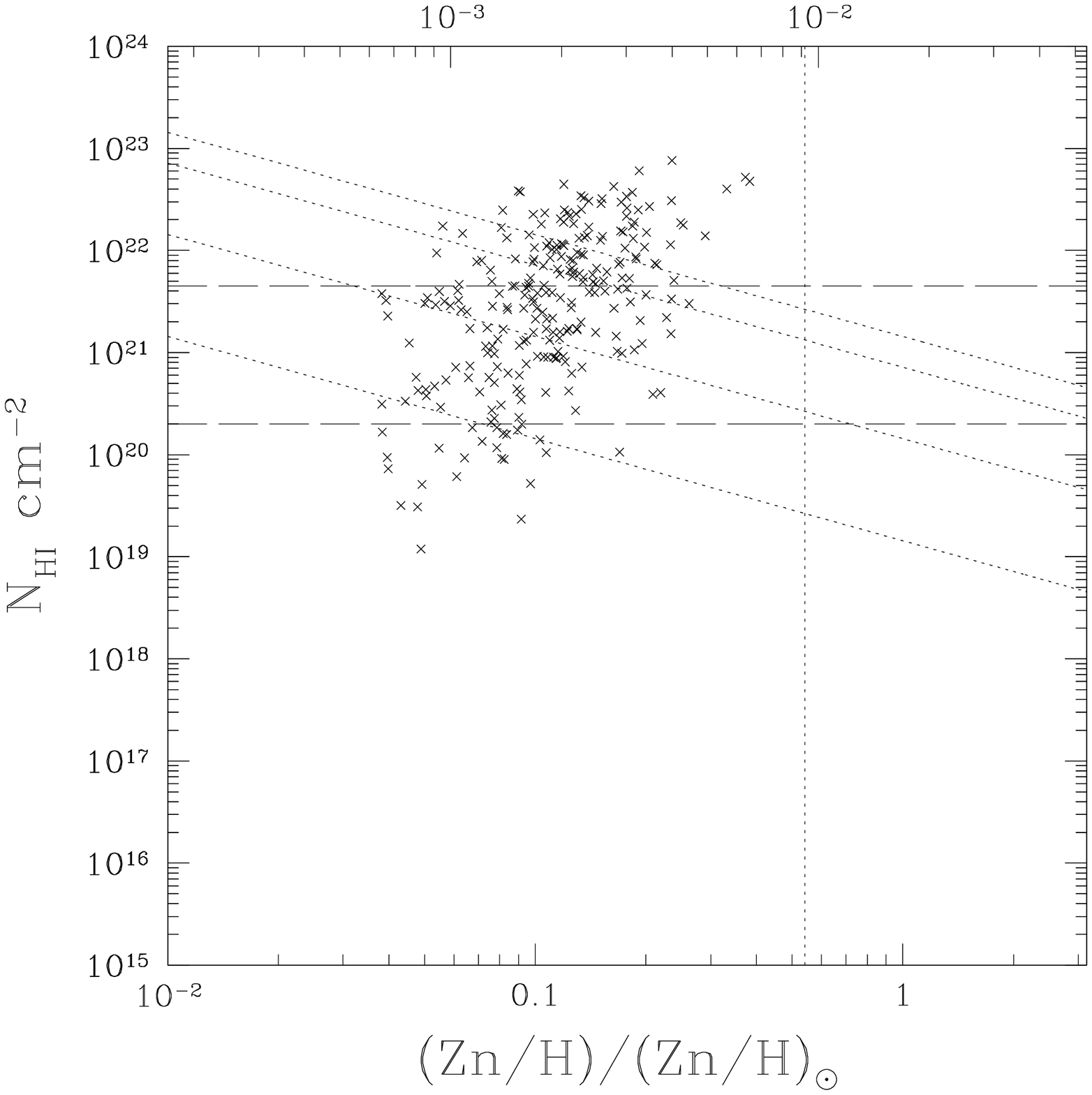,width=\widthxxyy,clip=}
&
\psfig{file=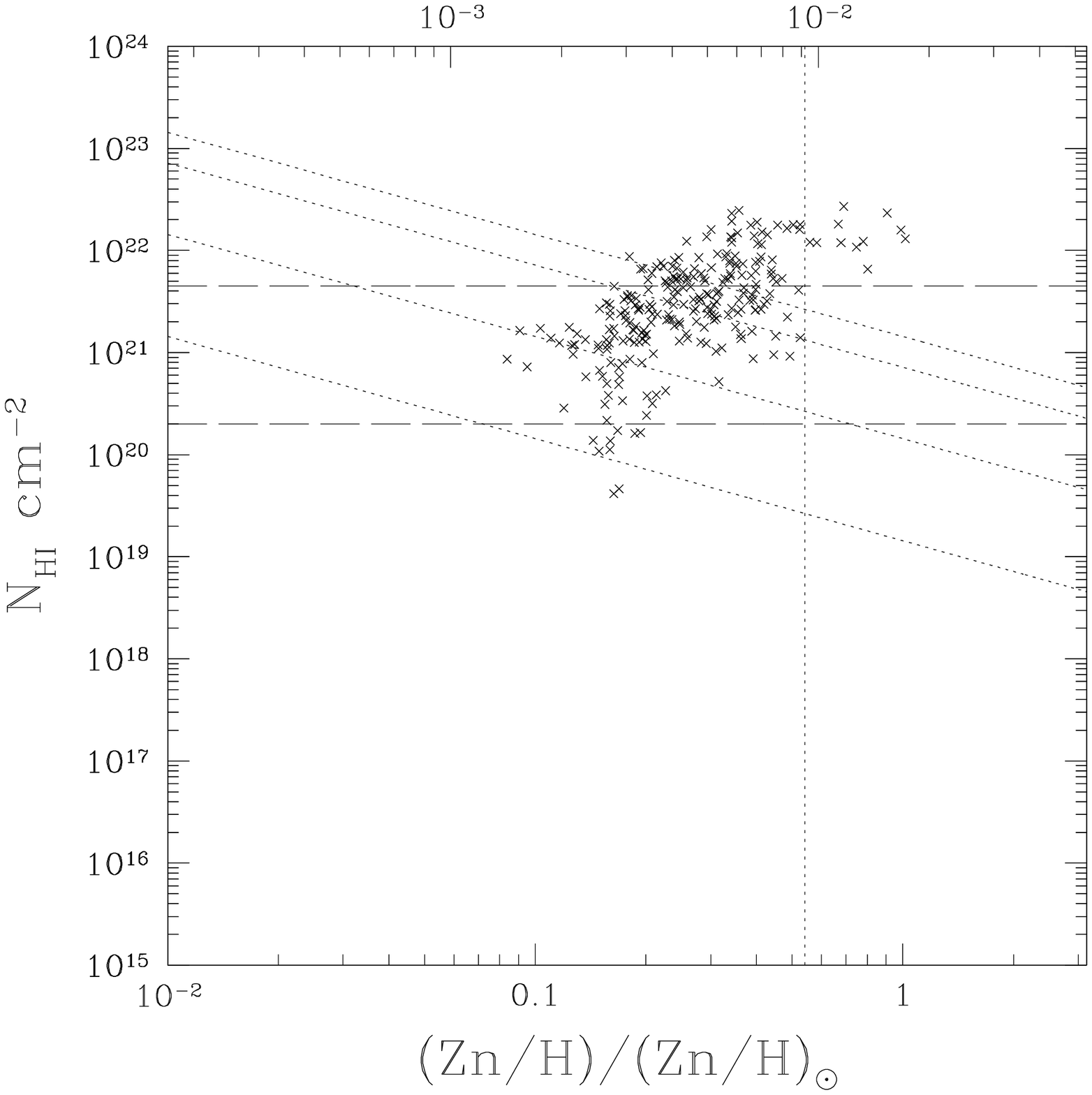,width=\widthxxyy,clip=}
\\
\psfig{file=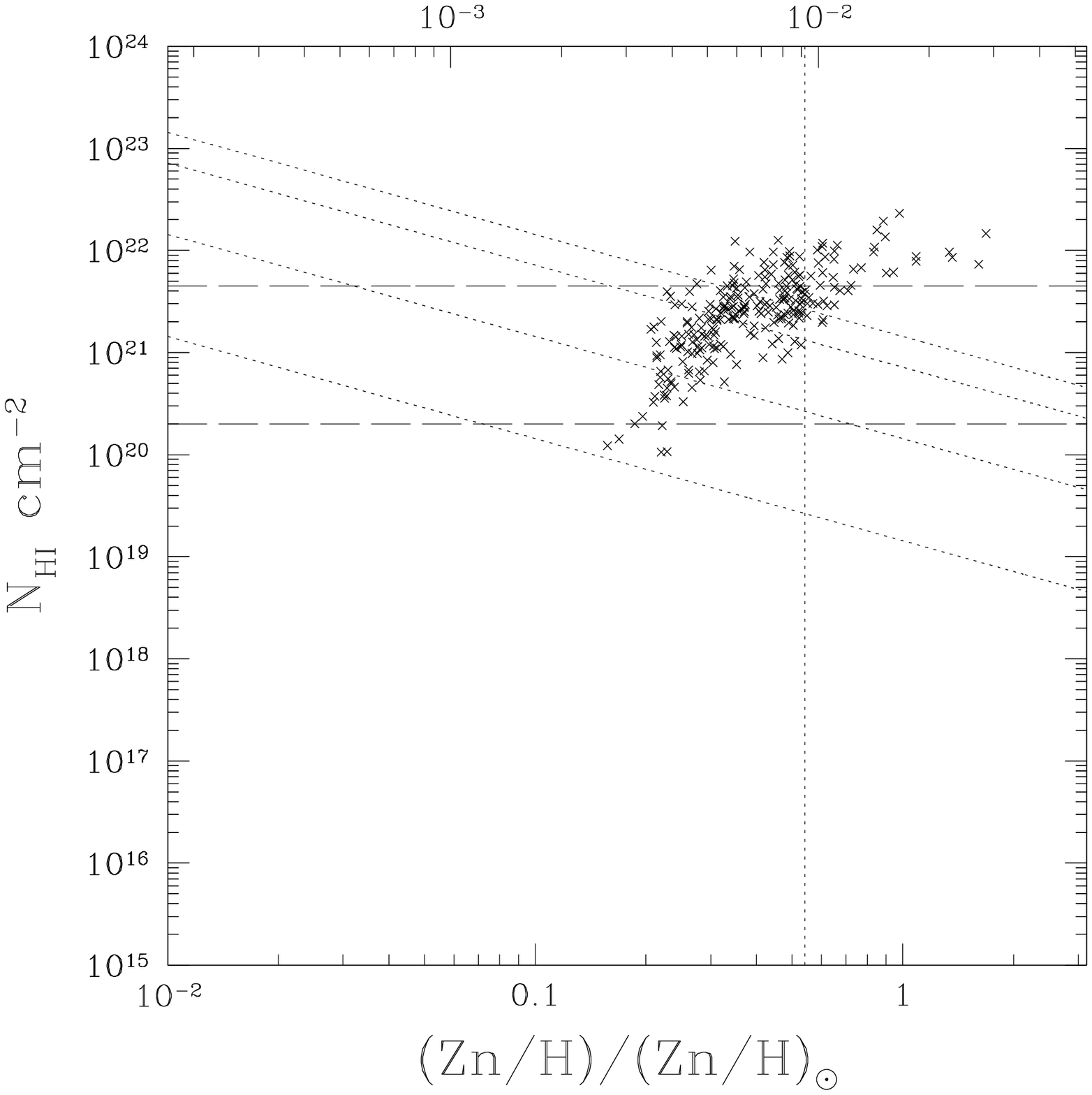,width=\widthxxyy,clip=}
&
\psfig{file=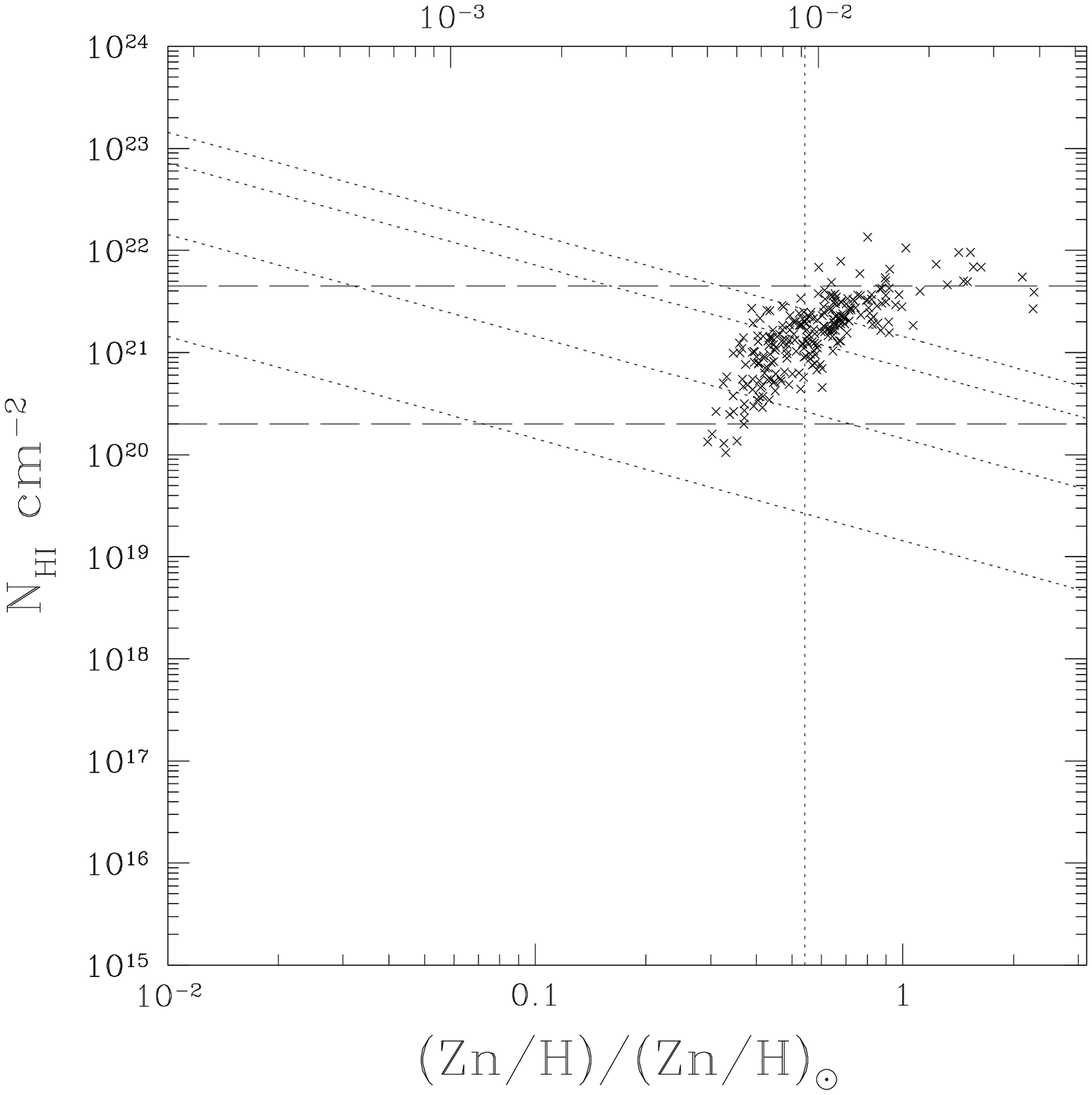,width=\widthxxyy,clip=}
\end{tabular}
\caption[]{Column densities and zinc abundances of
lines of sight through a model galaxy with 
mass $M=1\times 10^{12} \rm{M_{\odot}}$ and $\lambda=0.09$.}
\label{run5data}
\end{center}
\end{figure}

Figures \ref{run1data}, \ref{run2data}, \ref{run3data}, \ref{run4data} and \ref{run5data} show the results from the simulations.
The top left, top right, bottom left and bottom right panels of these figures show the simulations
at 1, 2, 3 and 5~Gyr respectively.

Figures \ref{run1data}, \ref{run2data} and \ref{run3data} show the results of varying $\lambda$ for fixed $M$, and figures \ref{run2data}, \ref{run4data} and \ref{run5data} show the results of varying $M$ for fixed $\lambda$.

In all cases the models fall within the box defined by the observations. This is consistent with the hypothesis that DLAs are indeed galaxy disks in different stages of evolution. At 1Gyr of evolution, heavy element abundances are small, and the sight lines occupy the left side of the diagram. As time increases, stars are formed, heavy elements produced, and the points move to the right. 

Although the models overlap with the observations, they occupy a greater region in the diagram.
As the sight lines move to the right of the diagram, for a given column density the optical depth increases. Therefore at early times it is possible to have a high column density sight line which is optically thin due to the fact that the metallicity is small. However, as time proceeds the models begin to have optical depths of $\tau=0.5$ and greater. 
The comparison between the models and observations can only be taken for $t=1$ Gyr and $t=2$ Gyr, since after this time the models have evolved past $z=1$. By $t=2$ Gyr, all of the models have a significant number of sight lines which have column densities greater than the observed upper limit. In addition, these lines lie above the $\tau=0.5$ line, and so have significant optical depth. Very few sight lines have evolved to the maximum observed metallicity by this time, which may explain the paucity of such systems in observations. 

\begin{figure}
\begin{center}
\begin{tabular}{@{}lr@{}}
\psfig{file=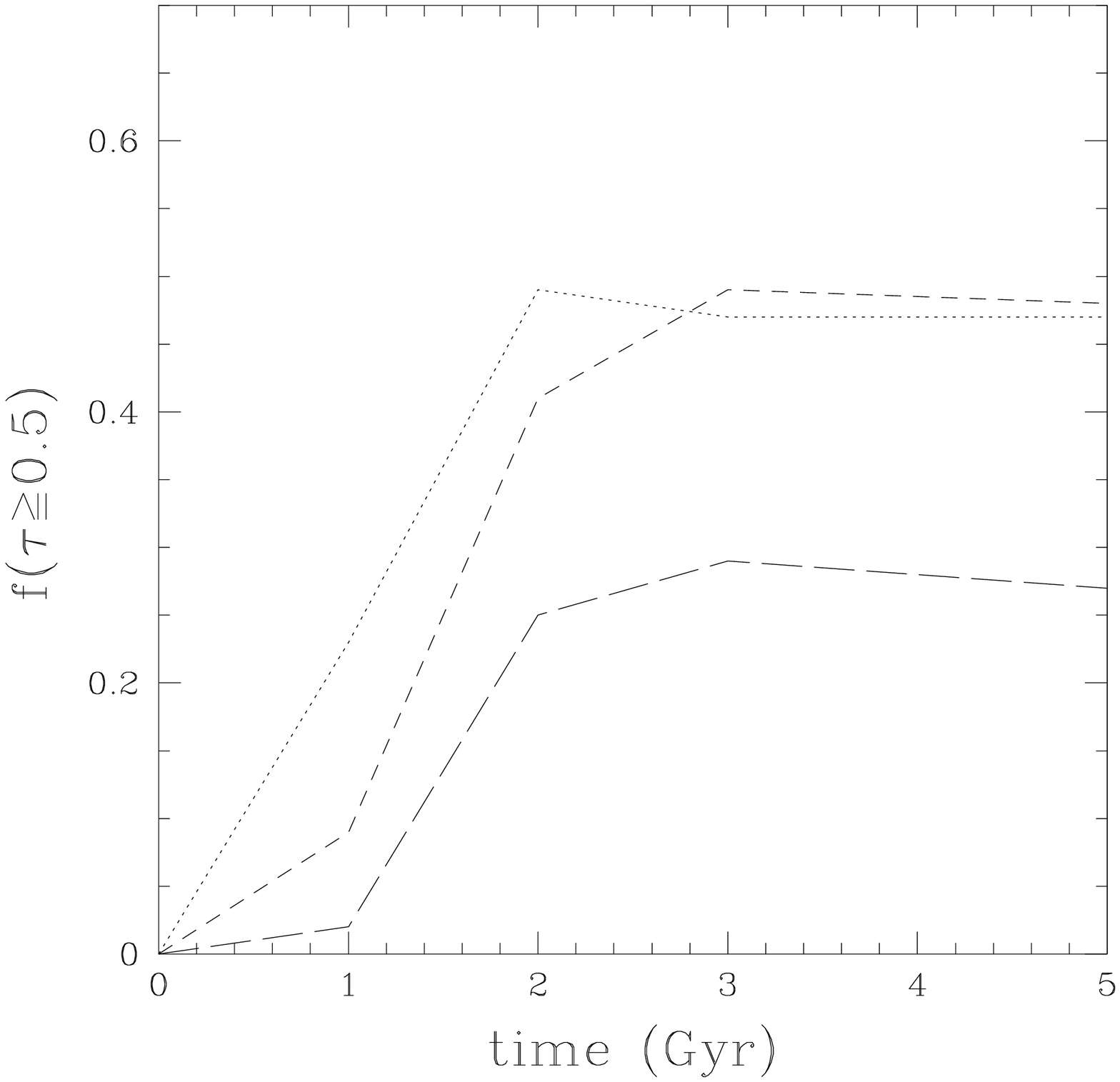,width=\widthxxyy,clip=}
&
\psfig{file=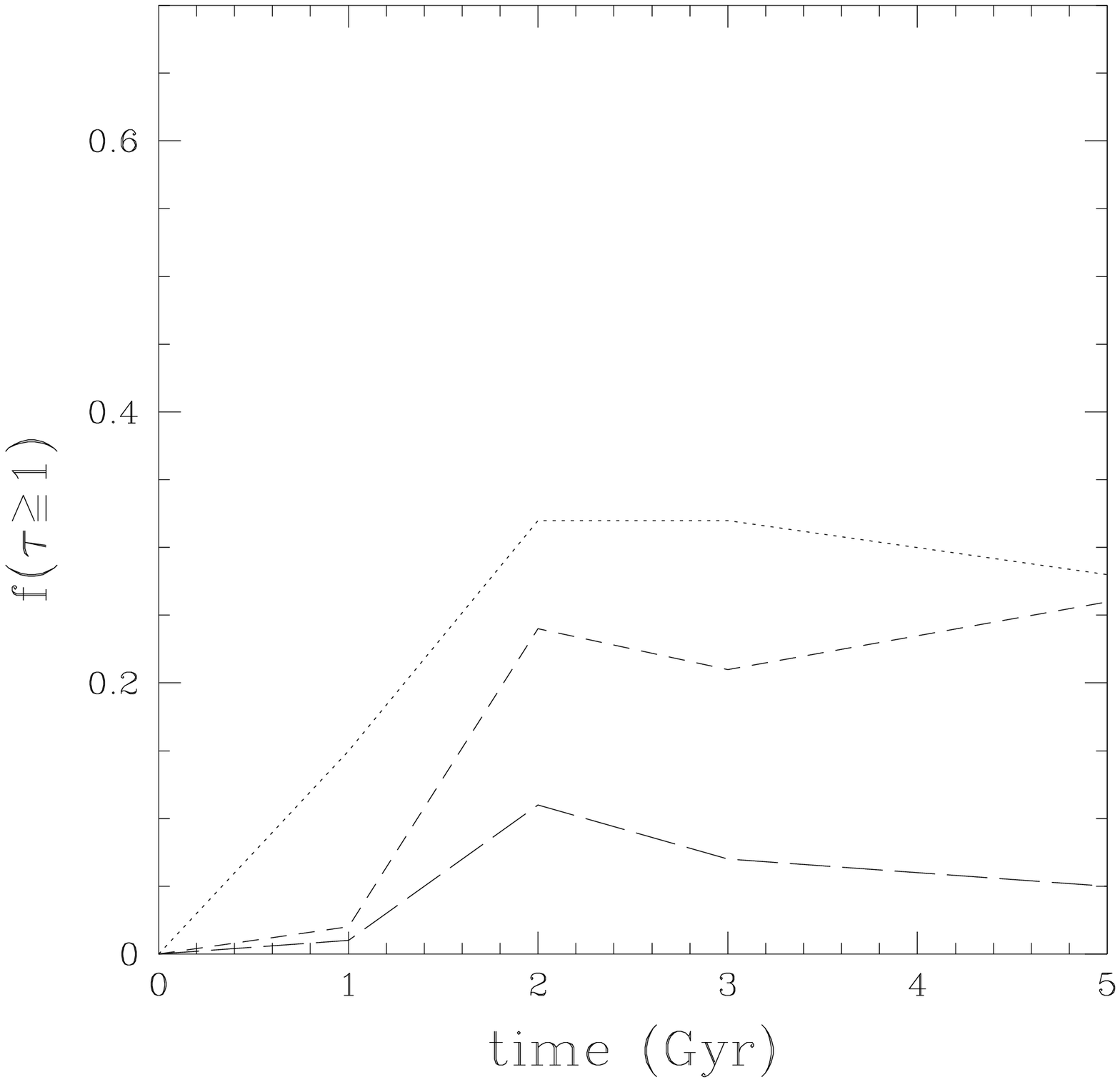,width=\widthxxyy,clip=}
\\

\psfig{file=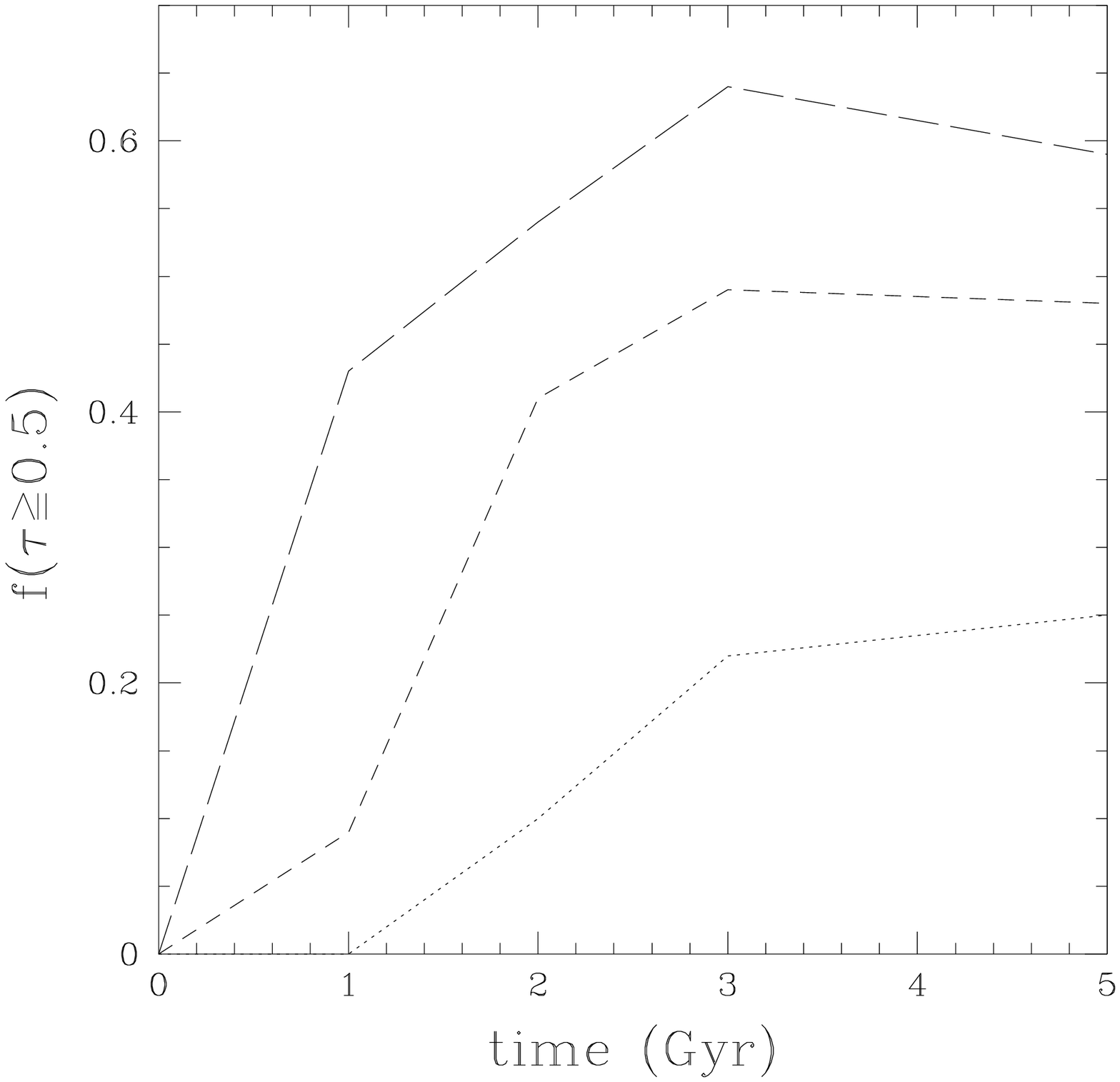,width=\widthxxyy,clip=}
&
\psfig{file=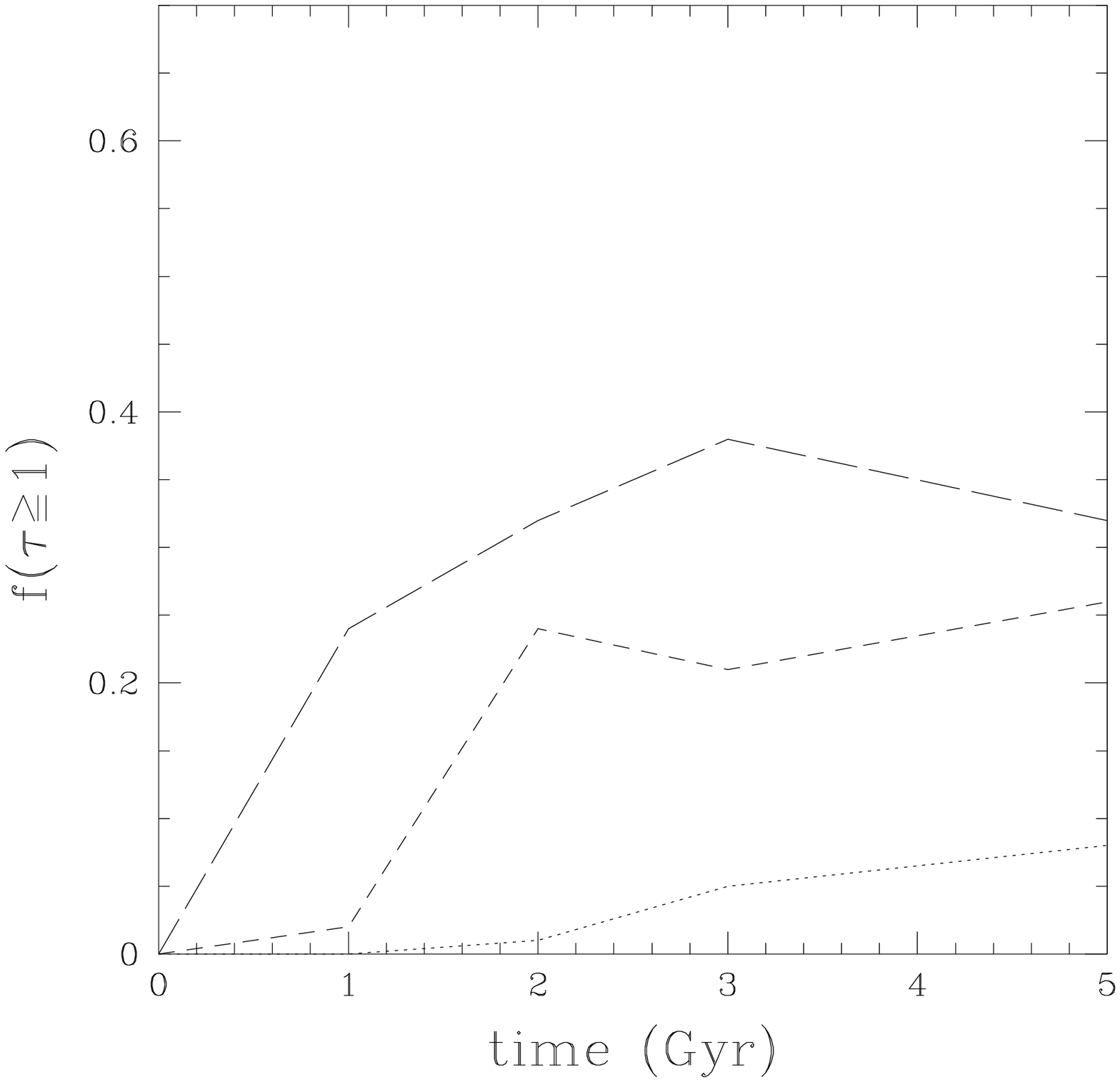,width=\widthxxyy,clip=}
\end{tabular}
\caption[Fraction of sight lines with optical depth greater than 0.5 and 1.0. The top two graphs show the variation with $\lambda$ for fixed $M$. Here the dotted line has $\lambda=0.06$, the short--dashed line has $\lambda=0.09$ and the long--dashed line has $\lambda=0.12$. In each case the mass is
$M=5 \times 10^{11} \rm{M_{\odot}}$]
{Fraction of sight lines with optical depth greater than 0.5 (left figures)  and 1.0 (right figures). The top two graphs show the variation with $\lambda$ for fixed $M=5 \times 10^{11} \rm{M_{\odot}}$. Here the dotted line has $\lambda=0.06$, the short--dashed line has $\lambda=0.09$ and the long--dashed line has $\lambda=0.12$. In the bottom two graphs we show the variation with $M$ for fixed $\lambda$. Here the dotted line has $M=2.5 \times 10^{11} \rm{M_{\odot}}$,  the short--dashed line has ,$M=5 \times 10^{11} \rm{M_{\odot}}$,  the $\lambda=0.09$ and the long--dashed line has $M=1 \times 10^{12} \rm{M_{\odot}}$.}
\label{fractions}
\end{center}
\end{figure}

Figure \ref{fractions} shows the fraction of sight lines which have an optical depth greater than 0.5 and 1.0, as a function of time, for all of the models.
This figure shows that in all cases, the number of sight lines with optical depths greater than $\tau=0.5$ and $\tau=1.0$ grows sharply during the first 2 Gyr of evolution, and the number remains approximately constant thereafter. This is because most of the heavy element synthesis occurs in the first 2 Gyr of the simulation, i.e. the period in which the optical depth grows coincides with the redshift space which the observations inhabit, and it is valid to compare the two.

Figure \ref{fractions} shows that the models produce galaxies which have $\sim$20 - 60\% of the sight lines through them with an optical depth greater than 0.5 and $\sim$10 - 30\% greater than 1.0.
Figure \ref{fractions} shows that for a fixed mass, smaller $\lambda$ produces more lines with higher optical depths. This is to be expected since a system with less angular momentum collapses down to higher gas densities, and so produces more metals.
This figure also shows that for a fixed $\lambda$, larger $M$ produces higher optical depths. This is again to be expected since the more massive systems have higher gas densities, and so produce more metals.

\begin{table}
\begin{center}
\begin{tabular}{|c|c|c|c|} \hline
 t/Gyr    &  $M=2.5 \times 10^{11} \rm{M_{\odot}}$ &  $M=5 \times 10^{11} \rm{M_{\odot}}$ &  $M=1 \times 10^{12} \rm{M_{\odot}}$ \\ \hline
  0       &       4.1                              &          5.4                         &       7.1                            \\
  1       &       1.9                              &          2.2                         &       2.4                            \\ 
  2       &       1.2                              &          1.3                         &       1.4                            \\
  3       &       0.8                              &          0.8                         &       0.9                            \\
  5       &       0.3                              &          0.4                         &       0.4                            \\ \hline
 
\end{tabular}
\caption[]{The approximate redshift reached by the models as a function of simulation time elapsed.}
\label{zoft}
\end{center}
\end{table}

Since the sight lines are uniformly distributed throughout the galaxy, and each one
represents the same amount of surface area of the galaxy disk, the fraction of the
total number of DLA sight lines which have a given optical depth is
the same as the fraction of the surface area of the galaxy with that optical depth.
Therefore, for the range of parameters considered, we can
produce galaxy disks which have 20--60\% of their surface area having $\tau \geq 0.5$, and 10--30\% with $\tau \geq 1.0$.

We agree with the suggestion of \cite {PF95} that a galaxy disk with a
significant optical depth could obscure the light from the background quasar to the point that it becomes unobservable.
This would mean that the absorber would not make it into any observational DLA sample,
and could explain why the high column density,
high metallicity systems are not present in figure \ref{dla_obs}.
Calculations by \cite{Baker00} indicate
that an optical depth of $\tau=0.5$ could suffice to produce this selection effect.

\subsubsection{Inclination Effects}
 
\begin{figure}
\begin{center}
\begin{tabular}{@{}lr@{}}
\psfig{file=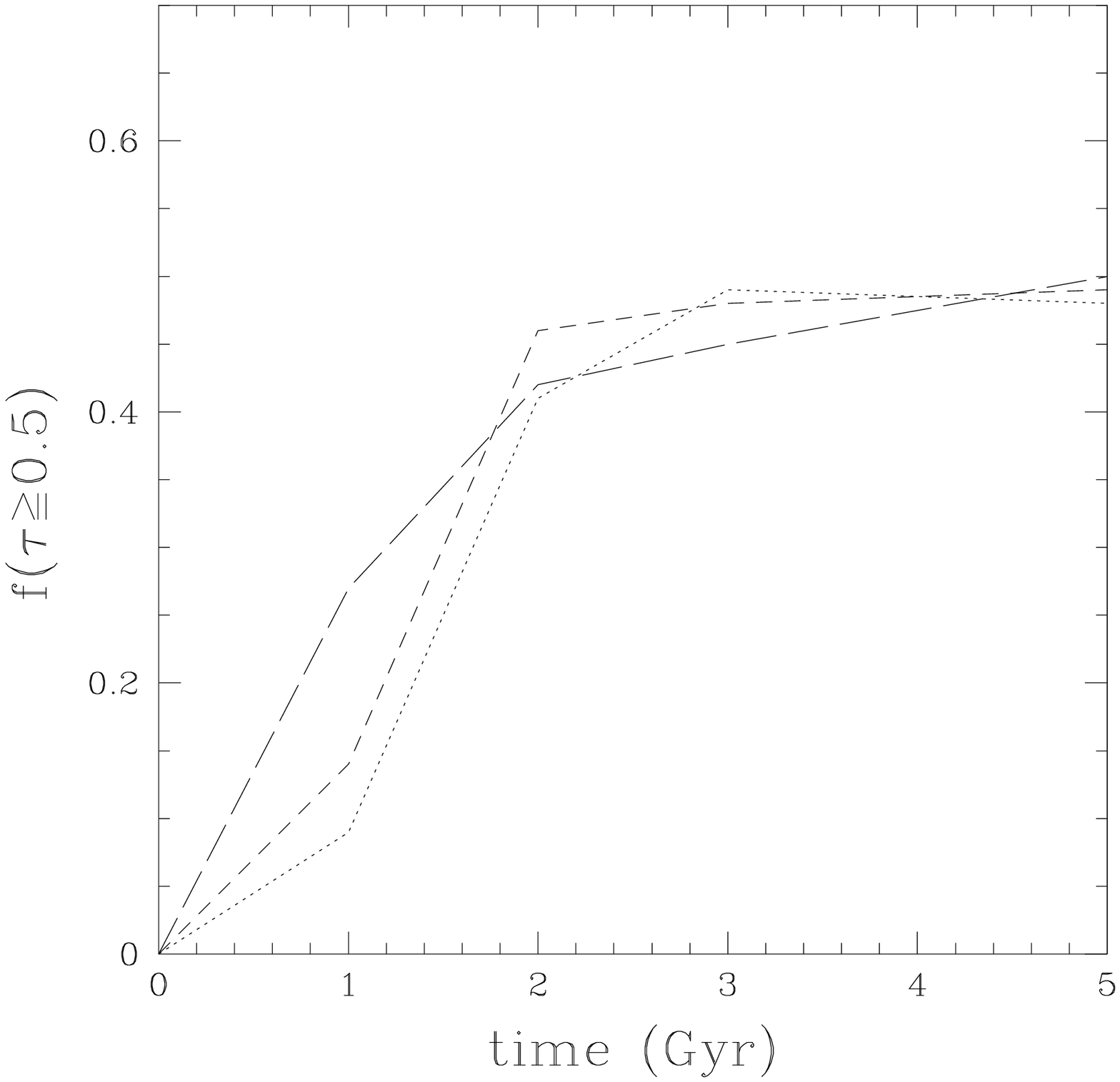,width=\widthxxyy,clip=}
&
\psfig{file=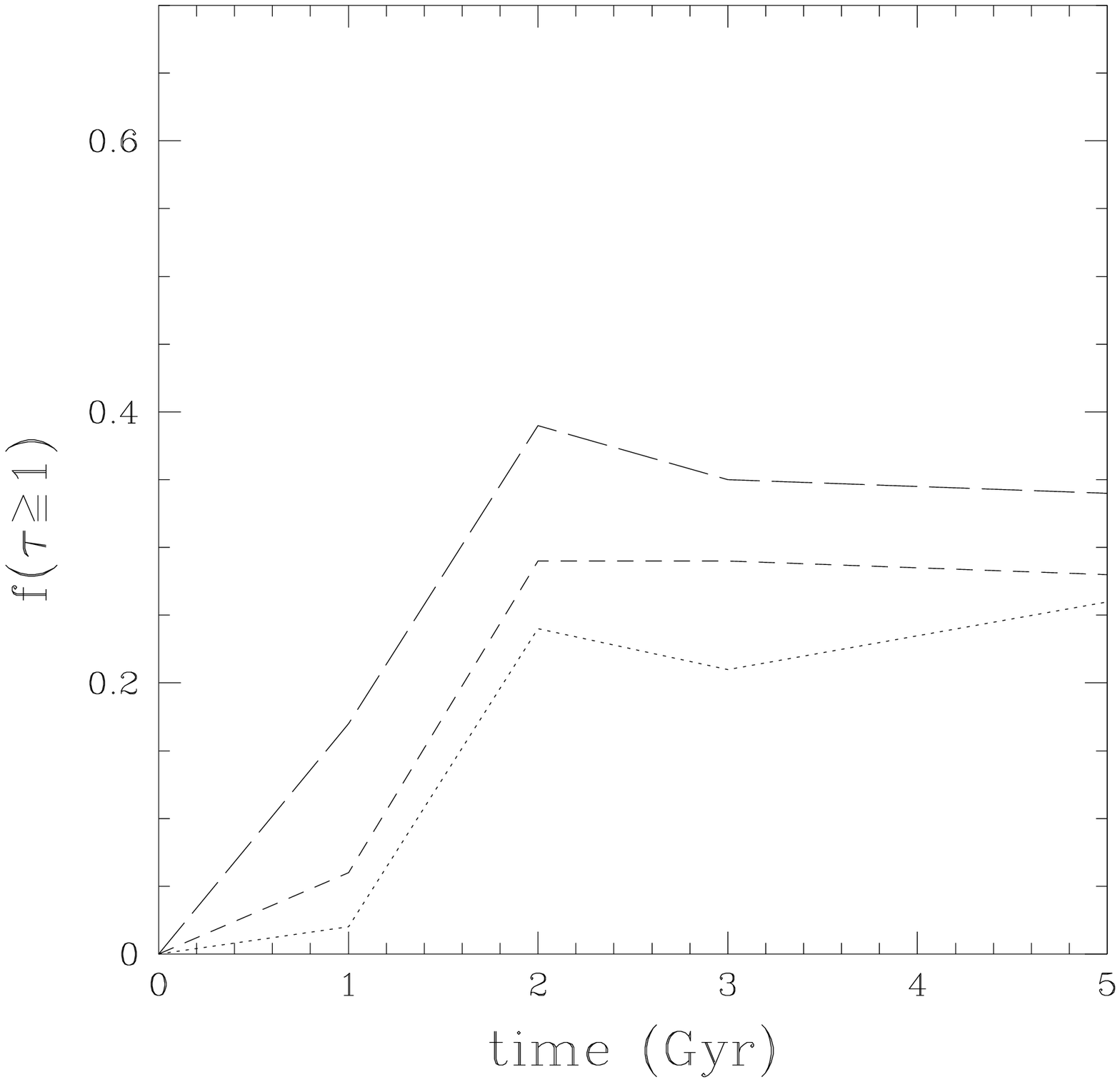,width=\widthxxyy,clip=}
\end{tabular}
\caption[ ]
{Fraction of sight lines with optical depth greater than 0.5 (left figure)  and 1.0 (right figure). The graphs show the variation with inclination angle $i$ for fixed $M=5 \times 10^{11} \rm{M_{\odot}}$. Here the dotted lines have $i=0^\circ$, the short--dashed lines have $i=60^\circ$ and the long--dashed lines have $i=80^\circ$.}
\label{fractionswiththeta}
\end{center}
\end{figure}

To investigate the effect of inclination upon our results, we have repeated the above analysis for one of our models with two additional inclination angles of $60^\circ$ and $80^\circ$. The results of this experiment are shown in figure \ref{fractionswiththeta}.

Figure \ref{fractionswiththeta} shows that the angle of inclination has a significant effect upon the sight lines which have an optical depth $\tau \geq 1.0$, increasing their fraction from $\sim 25\%$ for $i=0^\circ$ to $\sim 35\%$ for $i=80^\circ$.
Since the distribution function of inclination angles is a monotonically increasing function of the angle of inclination, one is more likely to observe highly inclined systems, and hence more likely to encounter higher optical depths.
Therefore the figures for the face--on models represent lower bounds on the fraction of the disk which has $\tau \geq 0.5$ or $\tau \geq 1$.

\section{Conclusions}
\label{conclusions}

We conclude that if the models presented here are representative of the observed DLA systems, a significant fraction of the sight lines through such objects will pass through regions with column densities in excess of the observed upper limit of $4.5 \times 10^{21}\,\,\rm{HIcm^{-2}}$.

If we assume that half of the mass which is in the form of heavy metals
is in the form of dust, then we predict
that in order for the simulations to produce such a sight line, the optical depth along that line would be
$\tau \geq 0.5$. \cite {PF95} have suggested that this may cause the light
from the background quasar to be sufficiently dimmed so that the quasar would drop out of the
observational sample, thus making the galaxy itself invisible in absorption.

By putting many sight lines through different parts of model galaxies we have
been able to calculate the fraction of the surface area of the disks which have an optical depth
greater than $\tau=0.5$ and $\tau=1$. For our face--on models we find that $\sim 20 - 60\%$ has $\tau \geq 0.5$ and $10 - 30\%$ has $\tau \geq 1$. If we increase the angle of inclination from $0^\circ$ to $80^\circ$ then the fraction of the disk which has $\tau \geq 1$ increases to $\sim 40\%$. Since highly--inclined systems are more likely to be observed, our face--on figures are lower bounds.

In addition, the most optically--thick region will typically be the inner part of the galaxy, and so the
observer may be in danger of drawing conclusions about the properties of these systems, based
only on observations of the \emph{outer} parts.

In conclusion, we support the hypothesis that the observed upper limit to the column density of damped Lyman alpha systems is a selection effect caused by obscuration by dust.
This implies that there could be significant numbers of high--metallicity, high redshift absorption systems as yet unobserved due to obscuration of their background QSO.

\section*{Acknowledgments}
We thank the anonymous referee for helpful comments which improved the presentation of this paper.

\bibliography{dave}
\bibliographystyle{mn2e}

\bsp

\label{lastpage}

\end{document}